\documentclass[letterpaper,twoside,twocolumn,english,prl]{revtex4-1}
\usepackage[T1]{fontenc}
\usepackage[latin9]{inputenc}
\usepackage{amsmath}
\usepackage{graphicx}
\usepackage{esint}

\makeatletter

\usepackage{xcolor}

\usepackage[]{ulem}

\usepackage{babel}

\makeatother

\usepackage{babel}
\begin{document}
\title{Record dynamics of evolving metastable systems: theory and applications}
\author{Paolo Sibani$^{1}$ Stefan Boettcher$^{2}$ and Henrik Jeldtoft Jensen$^{3,4}$}
\affiliation{$^{1}$FKF, University of Southern Denmark, Campusvej 55, DK5230,
Odense M, Denmark~~\\
 $^{2}$Department of Physics, Emory University, Atlanta, GA 30322,
USA~\\
 $^{3}$Centre for Complexity Science and Department of Mathematics,
Imperial College London, South Kensington Campus, SW7 2AZ, UK~\\
 $^{4}$Institute of Innovative Research, Tokyo Institute of Technology,
4259, Nagatsuta-cho, Yokohama 226-8502, Japan}
\begin{abstract}
Record Dynamics (RD) deals with complex systems evolving through a
sequence of metastable stages. These are macroscopically distinguishable
and appear stationary, except for the sudden and rapid changes, called
quakes, which induce the transitions from one stage to the next. This
phenomenology is well known in physics as ``physical aging'', but
from the vantage point of RD the evolution of a class of systems of
physical, biological and cultural origin is rooted in a hierarchically
structured configuration space and can therefore be analyzed by similar
statistical tools. This colloquium paper strives to present in a coherent
fashion methods and ideas that have gradually evolved over time. To
this end, it first describes the differences and similarities between
RD and two widespread paradigms of complex dynamics, Self Organized
Criticality and Continuous Time Random Walks. It then outlines the
Poissonian nature of records events in white noise time series, and
connects it to the statistics of quakes in metastable hierarchical
systems, arguing that the relaxation effects of quakes can generally
be described by power laws unrelated to criticality. Several different
applications of RD have been developed over the years. Some of these
are described, showing the basic RD hypothesis, the log time homogeneity
of quake dynamics, can be empirically verified in a given context.
The discussion summarizes the paper and briefly mentions applications
not discussed in detail. Finally, the outlook points to possible improvements
and to new areas of research where RD could be of use.
\end{abstract}
\maketitle

\section{\;\; Introduction}

Simply stated, the founding axiom of equilibrium statistical physics
is `all micro-states are equally probable in a thermally isolated
system'. Later coarse-graining criteria are of less sweeping generality.
Mori-Zwanzig projections, for example, aim to eliminate fast variables
from a set of interacting degrees of freedom (DoF), while diffusion
offers a probabilistic description of the configuration space movement
of particles subject to external white noise. In the same configuration
space, but describing a non-equilibrium process, Continuous Time Random
Walks~\cite{Kenkre73,Sher75} (CTRW) jump through a sequence of `traps',
each representing an attractor of the system's noiseless dynamics.
Importantly, choosing the Probability Density Function (PDF) of the
jump times to be a power-law allows one to describe e.g. sub-diffusive
behavior. While power laws are telltale signs of complex structure,
present, e.g., at the critical temperature of systems undergoing a
second order phase transition, their microscopic origin is not a focus
point of CTRW. Instead, they are assumed apriori, in form of the PDF.
Self Organized Criticality~\cite{Bak87,Jensen98book} posits that
a seemingly similar behavior is shared by certain systems which spontaneously
organize into a critical-like state, with no tunable external parameters
but while being slowly driven. Finally, diffusion in hierarchically
organized spaces~\cite{Grossmann85,Hoffmann88,Sibani89} generates
power-law relaxation laws with no reference to equilibrium properties,
e.g., critical points. In summary, there is a wide range of non-equilibrium
behaviors in systems with many interacting degrees of freedom, for
which there are often competing statistical descriptions that all
rely on simplifying assumptions and heuristic coarse-graining tools.

{ {blue}The statistics of extreme events is a subject with a long tradition. and many 
applications.
Gumbel~\cite{Gumbel58} famously studied the distribution of the largest out of $M$ independent identically distributed random  variables (i.i.d.). 
This statistics of extremes finds its place in standard text books, such as~\cite{Feller66}  and~\cite{Resnick87}, leading to ongoing
 studies into so-called "extreme-value statistics" and their universality classes~\cite{Bouchaud97,Majumdar08}.
  In~\cite{Krug07} the focus on the average and variance of the number of  records in time series of i.i.d.
  deviates whose distribution changes in time. The  properties of the record series have thus implications 
  regarding the series from 
  which the records are extracted. E.g. 
  deviations from the expected, independent production of records in a time series
   can be used to analyze hidden correlations. This   has been applied to  sports and climate data, or evolution~\cite{Sire09,Redner06,Krug05}.
  Recent progress in record statistics and its applications is discussed in~\cite{Wergen13}. 
  }
  { {red}
  In Record Dynamics (RD), record fluctuations of a locally stationary signal, e.g. thermal noise,  trigger macroscopic changes in metastable systems.
  The magnitude of these changes is linked to configuration space properties, e.g. the hierarchical barrier structure of the energy landscape,
  rather than  the size of the records. We are therefore  interested in temporal properties,
   such as the PDF of the waiting time for the next record in time series of i.i.d., e.g.  energy fluctuations in thermalized systems.
   The number of records in such time series is, to a good approximation,
   (log)Poisson distributed, see Eq.~\eqref{logPoisson} anf Fig.~\ref{fig:Rstat}.
  }
  
RD treats metastable systems evolving through a
sequence of `punctuated equilibria', a term invented by S. J. Gould~\cite{G+E}
in a macroevolutionary context. As we have argued extensively~\cite{Sibani13},
the same term also describes situations lacking time-translational
invariance and is used here in this more general sense. Sudden events,
or `punctuations', which are clearly distinguishable from reversible
fluctuations, lead from one metastable (pseudo-equilibrium) state
to the next, and control all macroscopic changes. The waiting times
for these de-facto irreversible events, dubbed `quakes', all have
expectation values which are finite but increase monotonically as
the system `ages' out of its initial state. The statistics of quakes
provides a coarse-grained description of the multi-scale behavior
characteristic of complex  systems 
, including the presence of power
laws for one and two point moments in the range of parameters specifying
the system's  glassy state. The strong dynamical similarities
of very different metastable systems is highlighted here by treating
each of them with the same RD techniques. The reader is referred to
the literature for more detailed descriptions of the cases presented.

The rest of this paper is organized as follows: Section~\ref{Background}
relates RD to Self-Organized Criticality and to diffusion in hierarchical
spaces. Section~\ref{Math} summarizes mathematical assumptions and
predictions and describes the techniques used to extract the quakes
and their statistics from available data. Section~\ref{App} reviews
a number of applications to observational, experimental and 
simulational data,  and Section~\ref{Out} summarizes the presentation and 
mentions  possible RD  extensions.

Finally, the figure included come from  works published over many years.
The notation has evolved during this period,
and we hope in the 
 reader's  understanding regarding the consequent non-uniform
 notation, mainly  in the figures and their  captions

\section{\;\; Background}

\label{Background} The configuration space of a metastable system
comprises many ergodic components differing in terms of macroscopic
observables~\cite{Palmer82}. Reversible fluctuations occur within
the components, while transitions from one component to another require
crossing sizeable free energy barriers, which may differ for a transition
and its reverse. If, within a certain observation time, typical equilibrium
fluctuations can trigger a transition one way but not the other, this
transition is \emph{de facto} irreversible. Irreversible transitions
violate detailed balance and their introduction would preclude an
accurate description of near-equilibrium states. In contrast, \emph{de
facto} irreversible transitions become reversible over time. In a
system with a continuum of barrier heights, increasingly many barriers
are crossed reversibly and ergodic components grow in size as the
system evolves out of its initial state. This situation typically
occurs whenever the initial quench generates a \emph{marginally stable},
i.e., a metastable state far from equilibrium stabilized by rather
small barriers.

Self Organized Criticality (SOC) and hierarchical diffusion models~\cite{Sibani01,Hoffmann88,Sibani89}
focus on different aspects of the dynamics out of marginally stable
states. Specifically, SOC describes stationary fluctuations of \emph{driven}
systems, while RD is inherently non-stationary. Nevertheless, RD shares
conceptual aspects with both SOC and hierarchical diffusion, but unlike
them uses de-facto irreversibility to treat barrier crossings as a
log-Poisson process. This provides an empirically verifiable
coarse-graining tool leading to  analytical predictions.

Below, we briefly discuss key modeling properties of SOC and diffusion
in hierarchies with focus on their relation to RD. The statistics
of records is dealt with  in the next section. Importantly, experimental
or numerical data usually do not directly deliver a series of record
events. Instead anomalous events, the quakes, must be identified which
allegedly correspond to transitions to new ergodic components. The
final step is to verify that these quakes are described by record
statistics, in short that they constitute a Poisson process whose
average has a logarithmic time dependence. Several examples are given
in Section~\ref{App}.

\subsection{\;\; SOC \& RD}

Self Organized Criticality (SOC) is a coarse grained description of
slowly driven, spatially extended systems with a large number of interacting
degrees of freedom, e.g., famously, the sand pile model~\cite{Bak87}.
In a SOC envisaged sandpile, grains of sand are added, one at the
time, eventually leading to a metastable situation when the local
slope of the pile exceeds its angle of repose. At some later time,
an avalanche occurs to restore the stability, or, rather, the marginal
stability of the pile. SOC macroscopic dynamics is thus described
by avalanches, which expectedly have a broad distribution of sizes
and of inter-event times. The metastable states are highly susceptible,
similar to critical states, in that a small perturbation, i.e., adding
a grain of sand, can elicit a large response, like an avalanche.

This section will not dwell on the large literature generated by the
SOC paradigm, see~\cite{HenrikSOC} and references therein, but will
instead take one step back in time to highlight the origin of some
key ideas that SOC and RD do share. Our starting point, to which we
refer for more details, is a paper by Tang, Wiesenfeld, Bak, Coppersmith
and Littlewood~\cite{Tang87}, for short TWBCL, predating the sand
pile model~\cite{Bak87}, The TWBCL model considers a linear array
of displacements $y_{j}$ of $N$ identical masses, each elastically
coupled to its nearest neighbors and subject to friction. The array
moves in a sinusoidal potential, driven by a train of square wave
pulses. In the limit of weak elastic couplings and high amplitude
of the square wave pulses, the periodic solutions of the problem are
characterized by the curvature ${\mathbf{C}}(n)$ of the array ${\mathbf{y}}(n)$
at each (integer) time $n$ right before the $n+1$ pulse begins.
In a metastable state, ${\mathbf{C}}$ does not depend on $n$. As
TWBCL point out, all metastable states lie in an $N$-dimensional
hypercube, centered at the origin. A randomly chosen initial state
will move toward the origin, but will stick to the surface of the
hypercube, once it gets there. Importantly, initial configurations
sufficiently far from the surface invariably end sticking at the \emph{corners}
of the hypercube. The states picked are minimally stable, and are
akin to the metastable states of the sandpile. Since the dynamics
is purely deterministic, stable states in the interior of the hypercube
are never accessed. Adding white noise to the dynamics allows the
system to relax logarithmically toward the center~\cite{Sibani01}.

In terms of concepts laid out by TWBCL, the difference between SOC
and RD is that the former considers jumps between corner states, induced
by external driving but free of internal fluctuations, while the latter
describes the slow aging process which follows from random fluctuations
by coupling to a heat-bath or some other form of external noise. TWBCL
highlights the presence of hierarchies of attractors and that processes
not described by  equilibrium considerations are key elements
of the dynamics. Ref.~\cite{Sibani93a} on `noise adaptation' uses
a simple model with a hierarchy of attractors of different stability
to show how white noise of bounded amplitude selects marginally stable
attractors which cannot be escaped without infinitesimally increasing
the noise amplitude. If the latter does not have bounded support,
e.g. for Gaussian noise, this translates into an increasing stability
of marginally stable attractors, which is a feature of RD.

\subsection{\;\; Hierarchies \& RD}

The connection between complexity and hierarchies was emphasized long
ago by Herbert A. Simon~\cite{Simon62} in his seminal essay `The
Architecture of Complexity'. Simon defines a hierarchy as a system
composed of subsystems, each again composed of subsystems, etc , as
in a set of russian Matroska dolls. Many physical, biological and
social systems conform to this paradigm. In any case, a process initiated
within a sub-unit must, in the short run, remain confined to that
sub-unit, and only in the long run it is able to affect the whole
structure. The dynamics is then nearly \emph{decomposable}. Furthermore,
the hierarchical structure of the system implies that the propagation
of an initially localized perturbation is a decelerating multi-scaled
relaxation process.

As shown in Section~\ref{SG}, the hierarchical nature of complex
relaxation can already be inferred from cleverly designed experiments.
It can, however, also be directly ascertained from studying numerical
models: Using the lid method~\cite{Sibani99}, all microscopic configurations
of a discrete system whose energies lie between a local energy minimum
and a preset `lid' energy value are enumerated. The master equation
for the system can then be set up and solved numerically, obtaining
for each time $t$ the probability density $P(t,x|x_{0})$ of finding
the system in a micro-state $x$, given a sharp initial condition
at $x_{0}$. Two micro-states $x$ and $y$ are considered to be locally
equilibrated, if $\frac{P(t,x)}{P(t,y)}\approx\frac{\exp(E(x)/T)}{\exp(E(y)/T)}$,
where near equality is controlled by a small allowed deviation. All
micro-states can then be grouped into classes. Since the system will
eventually equilibrate, all states will eventually end up in the same
class, and a merging of different classes can be expected during the
time evolution of the system.

\begin{figure}
\vspace{-0.2cm}
 \hfill{}\includegraphics[width=0.5\textwidth]{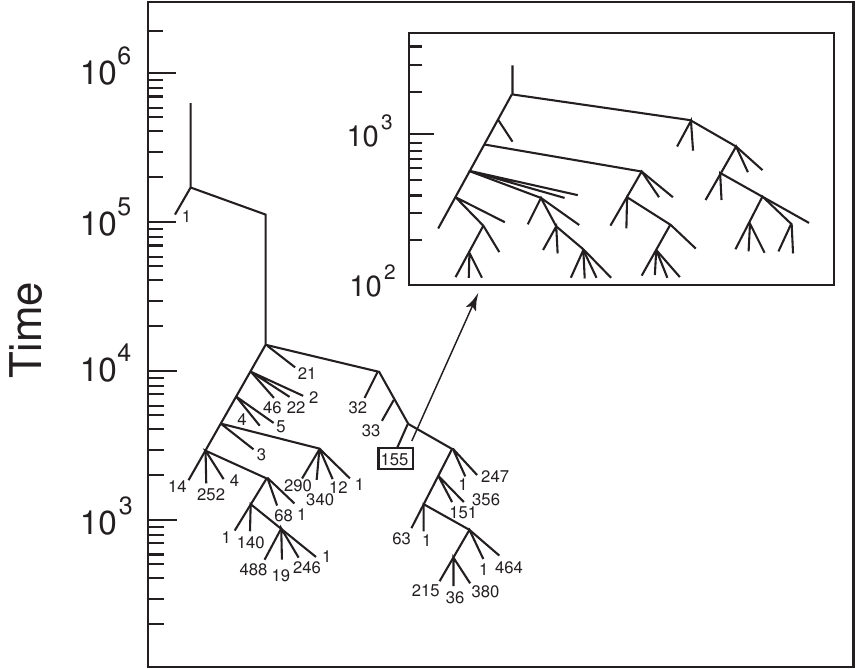}\hfill{}
\caption[Relaxation tree for the TSP]{{\small{}The figure, taken from Ref.~\cite{Sibani93} shows the
hierarchical structure of relaxation at fixed temperature in the state
space of a Traveling Salesman Problem. Each leaf of the tree represents
a set of states which are in local equilibrium with one another at
a certain level of approximation $\epsilon$. The insert shows how
a leaf splits into a whole tree when a smaller value of $\epsilon$
is chosen. The vertical axis of the tree is proportional to the logarithm
of time: thus the system undergoes a series of local equilibrations
(= merging of subtrees ) at times almost equally spaced on a logarithmic
scale.}}
\label{rel_tree}
\end{figure}

Figure \ref{rel_tree}, taken from~\cite{Sibani93}, shows what happens
when the procedure is applied to a small instance of the Traveling
Salesman Problem, where the tour length plays the role of energy.
Equilibration proceeds in stages, with different classes merging,
but never splitting again. In the resulting relaxation tree, branches
merge at times nearly equally spaced on a logarithmic scale.
Considering the thermally activated nature of the process, this indicates
that the energy barriers allowing  different quasi-equilibrium states
to  merge are equally spaced on a linear scale.

\begin{figure}
\hfill{}\includegraphics[bb=0bp 237.6699bp 784.08bp 564.466bp,clip,width=1\columnwidth]{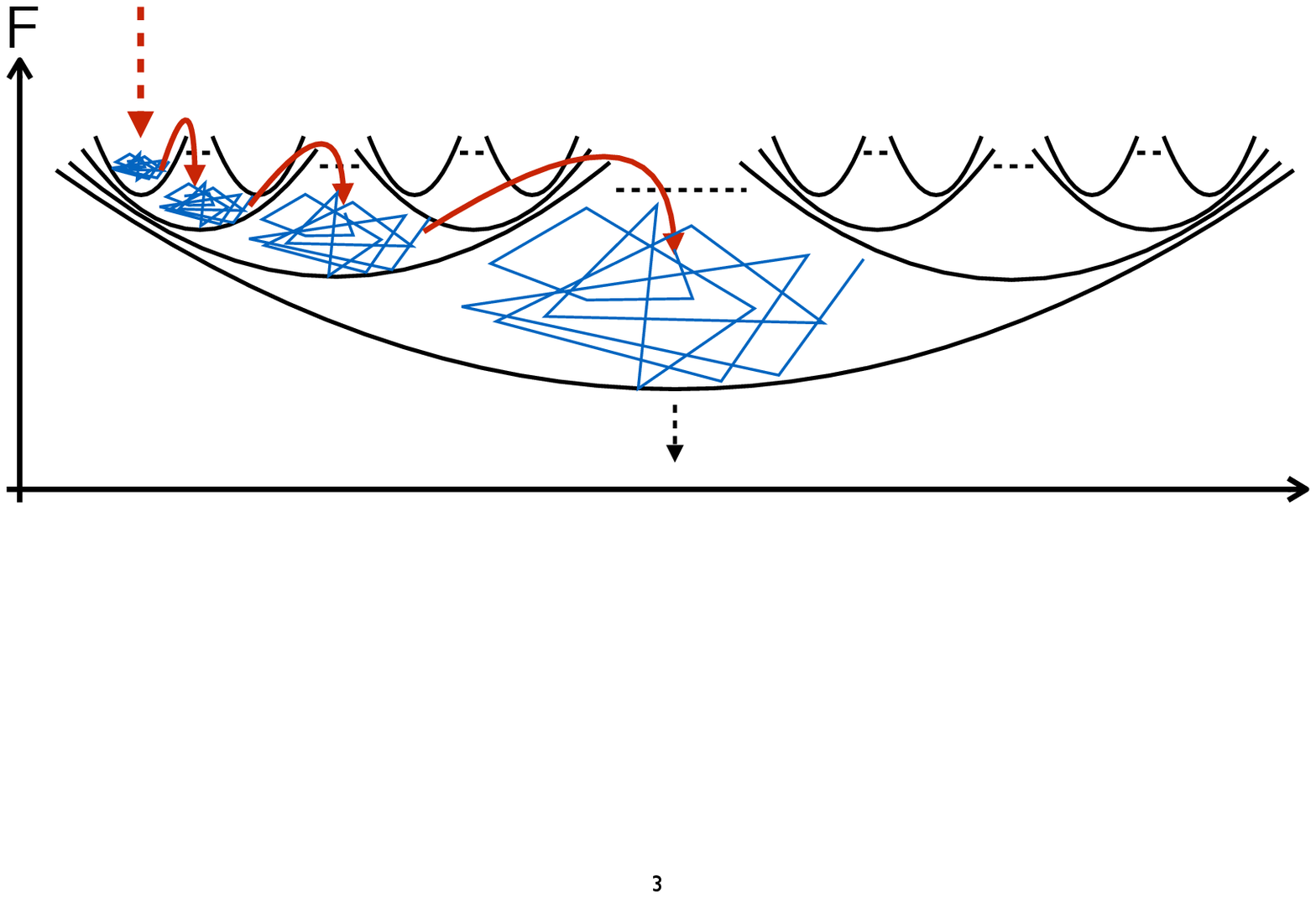}\hfill{}

\caption{{\small{}\label{fig:FLandscape} Sketch of the hierarchical (free-)energy
landscape of a complex system in RD, from Ref.~\cite{Robe16}, with
a typical trajectory of an aging dynamics (blue and red). With increasing
free energy $F$, local minima proliferate exponentially but also
become shallower. After a quench (red-dashed arrow), the dynamics
evolves through a sequence of quasi-equilibrium explorations (blue)
and intermittent, irreversible quakes over record barriers (red) that
access an exponentially expanding portion of configuration space (black-dashed
arrow).}}
\end{figure}

To describe thermal relaxation dynamics, the configuration space of
complex systems can be coarse-grained in a tree graph~\cite{Hoffmann88}
of the kind illustrated by Fig.~\ref{rel_tree} In RD terminology,
the stepwise process of relaxation in such a hierarchy is controlled
by a series of quakes, each associated to the crossing of a record
high barrier, and each giving access to a hitherto unexplored region
of configuration space. If the barriers are uniformly distributed
on the energy axis, the quakes are uniformly distributed on a logarithmic
time scale. The simplified picture sketched above neglects the discrete
nature of the energy barrier structure but ensures that quaking is,
as assumed by RD, a Poisson process whose average is proportional
to the logarithm of time, for short, a log-Poisson process.

An alternative description of the nature of the processes described
with RD is facilitated in terms of (free-)energy landscapes. As such
landscapes are widely used to conceptualize disordered materials,
it illustrates further the wide applicability of RD. A relaxing system
is characterized by growing domains, however slowly, of correlated
behavior between DoF. For domains to grow (in an otherwise finite
system), others have to disappear, a process referred to as ``coarsening''.
To overturn such a domain in a complex system requires crossing an
energy barrier. It is known~\cite{Shore92} that there are only two
generic scenarios: either barrier energies grow with the size of the
domain to be overturned, or they are independent. In the latter case,
average sizes of the remaining domains grow as a powerlaw with time,
$\sim t^{1/2}$, but if bigger domains cost ever more energy to overturn,
their growth is merely logarithmic with time, $\sim\ln t.$ While
such growth may be hard to observe on any experimental scale, it still
enforces the following feedback loop: The landscape of such a complex
system has a hierarchical structure in that, the lower energies (i.e.,
larger domains) are reached, the higher the barriers get, and thus,
larger fluctuations are needed to escape local minima~\cite{Robe16}.
Such a feedback is absent in the coarsening of, e.g., an Ising ferromagnet,
where barriers remain independent of domain size, thus, independent
of the depth within the landscape. (If entropic effects become relevant
~\cite{Shore92}, like for a structural glass, the same argeument holds
for free-energy barriers.)

Hence, RD presupposes three features that are well-established for
hierarchical energy landscapes: (1)\ Meta-stable states, and thus
their combined basin of attraction, proliferate exponentially for
increasing free energy~\cite{Bray1980,Stillinger84,Stillinger99,Heuer08};
(2)\ more-stable (lower-energy) states typically have higher barriers
against escaping their basin~\cite{Sibani93b,Heuer08,Shore92}; and
(3)\ higher jumps in energy make exponentially more configurations
accessible~\cite{Sibani89,BoSi}. Such a landscape is illustrated (as
a projection of a configuration space with a large number of DoF)
in Fig.~\ref{fig:FLandscape}. In a quench, the system almost certainly
gets stuck initially in one of the many shallow basin of high energy.
There, a small, random fluctuation already suffices to escape into
a larger basin containing many sub-basins, some of which feature local
minima of lower energy. But for reaching basins of ever lower energy,
gradually, higher fluctuations are required to escape. The gain in
stability acquired in any one of these escapes will most probable
only be \emph{marginal}, since spontaneously finding minima of dramatically
lower energy would be \emph{exponentially} unlikely. The motion in
and out of states less stable than the current basin only provides
reversible quasi-equilibrium fluctuations. For an irreversible quake,
typically, only a rare record fluctuation in the noise impinging on
the system will suffice~\cite{Sibani93a}, whether it is in a spin
glass~\cite{Crisanti04,Sibani05}, in a colloidal glass~\cite{Yunker09,Robe16,Tanaka17},
or even in a athermal (tapped) granular pile~\cite{GB20}. RD, hence, offers an analytic
in-road to coarse-grained descriptions that bridge the aging phenomenology.
This approximation supersedes the particular properties of the hierarchy
of barriers in a given system, e.g. small differences in energy or
density, or some microscopic details. As long as such a hierarchy
exists~\cite{Fischer08,Joh96,Liao2019,PSAA,Sibani89,Toninelli12,Scalliet19,Sibani93b,Charbonneau13},
i.e., the system is actually jammed, microscopic distinctions only
vary by an overall unit of time.

The examples presented in the following show how the toy pictures
just discussed must be nuanced. A hierarchical barrier structure captures
nevertheless metastability and is thus a prerequisite for RD to be
applicable.

\section{\;\; Mathematical backbone}

\label{Math} A stationary time series of e.g. measured temperature
values in a thermostated system, qualifies as white noise if successive
measurements are statistically independent, which can be achieved
by taking them sufficiently far apart. As will become apparent, the
statistics of records is independent of the mechanism generating the
noise. In particular it is independent of temperature in the case
of temperature records in thermal fluctuations.
{ {red} The universality of record statistics lies behind
the shared phenomenology of microscopically very different metastable 
systems which experience  punctuated equilibria. Below
 the 
mathematical properties of record statistics are highlighted which
can  be easily extracted  from time series of experimental,
observational or simulational data. 
 }

Consider a stationary series $S_{k},\;k=0,1\ldots$ of statistically
independent and identically distributed scalar random variables, so-called
\emph{white noise}. { {magenta} We assume that the distribution is not
 supported on  a finite set.}
 By definition, the first entry of the series is
a record. Other entries are records if and only if they are larger
than the previous record. Memory of each record is maintained by an
algorithm, and the issue of how record fluctuations can leave an indelible
mark in the evolution of a physical system does not arise in this
connection.

Record-sized entries within form a sub-series ${\mathbf{R}}\subset{\mathbf{S}}$
whose interesting statistical properties follow from simple arguments~\cite{Glick78,Sibani93a,Sibani98a,Nevzorov01}:
As it gets increasingly harder for an entry to qualify as the next
record, records appear at a decreasing rate and the sub-series of
record-sized entries is not stationary. The number of records in the
time interval $(1,t)$ is to a good approximation a Poisson process
with average $\ln t$. Equivalently, if the $k$'th record appears
at times $t_{k}$, the ratios $\ln(t_{k}/t_{k-1})$ are independent
random variables with an exponential distribution. The above properties
are derived using heuristic approximations, and their validity is
checked numerically.

\subsection{\;\; Record statistics in white noise}

\label{Records_in_noise}

Denote by $p_{n}(l,[0,m_{1}\ldots m_{n}])$ the probability that $n$
records are located at entries $0$, $m_{1},\ldots,m_{n}$ of the
series $\mathbf{S}$, and let $P_{n}(l)$ be the probability that
$n$ out of the $l+1$ entries are records, regardless of their location.
Clearly, in the case $n=1$, the first entry must be the largest.
As the largest entry can be anywhere with equal probability, 
\begin{equation}
P_{1}(l)\stackrel{{\rm def}}{=}p_{1}(l,[0])=\frac{1}{l+1},\quad l=0,1\ldots\infty,\label{one_record}
\end{equation}
is the probability of finding precisely one record in the first $l+1$
elements of the series. For two records, one occurring at $0$ and
the other at $m_{1}\le l$, we find 
\begin{equation}
p_{2}(l,[0,m_{1}])=\frac{1}{m_{1}}\frac{1}{l+1},\quad1\le m_{1}\le l.\label{two_records_joint}
\end{equation}
The first factor on the r.h.s. of the equation is the probability
that the zero' th element is the largest of the first $m_{1}$ entries,
and the second is the probability that the largest element of the
whole series be located at $m_{1}$. The probability that exactly
two out of $l+1$ elements are records is then 
\begin{equation}
P_{2}(l)=\frac{1}{l+1}\sum_{m_{1}=1}^{l}\frac{1}{m_{1}}=\frac{H_{l}}{l+1}\label{Eq: two_records_joint_also}
\end{equation}
where the \emph{harmonic number} $H_{l}=\sum_{k=1}^{l}1/k$ satisfies
\begin{equation}
H_{l}=\ln(l+1)+\gamma+{\cal O}(l^{-1})\label{Harmonic number}
\end{equation}
with $\gamma=0.57721\ldots$ being the Euler-Mascheroni constant.
Turning now to an arbitrary $n$, we find, by the same arguments,
\begin{eqnarray}
 &  & p_{n}(l,[0,m_{1},m_{2},\ldots m_{n-1}])=\frac{1}{l+1}\prod_{k=1}^{n-1}\frac{1}{m_{k}},\\
 &  & {\rm with};\quad1\le m_{1};\quad m_{k-1}<m_{k};\quad m_{n-1}\le l.\label{n_records_joint}
\end{eqnarray}
To obtain $P_{n}(l)$, the distribution $p_{n}(l,[0,m_{1},m_{2},\ldots m_{n-1}])$
must be summed over all possible values of $m_{1},m_{2}\ldots m_{n-1}$.
Unfortunately, a closed form expression can only be obtained in the
continuum limit, where all sums turn into integrals.

To carry out the needed approximation, assume that $\mathbf{S}$ is
obtained by sampling a stationary signal at regular intervals of duration
$\delta t$. Furthermore, let $t_{k}\stackrel{{\rm def}}{=}m_{k}\delta t$
and $t\stackrel{{\rm def}}{=}l\delta t$ be the time of occurrence
of the $k$'th record in the series, and the total observation time,
respectively. Noting that $m_{k}=t_{k}/\delta t$, and taking the
limit $\delta t\rightarrow0$, we find 
\begin{equation}
P_{n}(t)=\frac{1}{t}\int_{t_{n-2}}^{t}\frac{dt_{n-1}}{t_{n-1}}\prod_{k=n-2}^{2}\int_{t_{k-1}}^{t_{k+1}}\frac{dt_{k}}{t_{k}}\int_{1}^{t_{2}}\frac{dt_{1}}{t_{1}}
\end{equation}
As the integrand in the above expression is a symmetric function of
the arguments $t_{1},\ldots,t_{n-1}$, any permutation of the arguments
does not change the integral. Summing over all permutations of the
order $n-1$ variables and dividing by $(n-1)!$ leaves the expression
unchanged and yields 
\begin{eqnarray}
P_{n}(t) & = & \frac{1}{t}\frac{1}{(n-1)!}\left(\int_{1}^{t}\frac{dz}{z}\right)^{n-1}\nonumber \\
 & = & \frac{1}{t}\frac{\log(t)^{n}}{n!},\quad n=0,1,2,\ldots\infty
 \label{logPoisson}
\end{eqnarray}
In the last expression, which can be recognized as a Poisson distribution
with expectation value 
\begin{equation}
\mu_{n}(t)=\log(t),
\label{mean_var}
\end{equation}
the convention that the first entry is always a record is abandoned
for convenience. As well known, the variance of the process equals
its mean, $\sigma^{2}(t)=\mu_{n}(t)$.

Clearly, the number of records falling between times $t_{w}$ and
$t>t_{w}$ is the difference of two Poisson process, and hence itself
a Poisson process with expectation $\mu_{n}(t_{w},t)=\ln(t)-\ln(t_{w})=\ln(t/t_{w})$.
The average number of records per unit of time decays as 
\begin{equation}
\frac{d\mu_{n}(t)}{dt}\stackrel{{\rm def}}{=}\frac{1}{t}.\label{oneovert}
\end{equation}
Importantly, the logarithmic rate of events is constant,
and the transformation $t\rightarrow\ln t$ renders the series $\mathbf{R}$
memoryless and translationally invariant. 

A Poisson process is uniquely characterized by the exponential PDF
of the waiting time to the next event. In our case, Eq. \eqref{logPoisson}
implies that $\ln t$ plays the role of time variable, which brings
our focus on the logarithmic waiting time between successive events,
the $k$'th waiting time being $\Delta_{t_{k}}=\ln t_{k}-\ln t_{k-1}=\ln(t_{k+1}/t_{k}),\;k\ge1$.
The $\Delta_{t_{k}}$ are independent of $k$ and exponentially distributed
with unit average. Their exponential PDF is thus
\begin{equation}
F_{\Delta{\ln t}}(x)=\mbox{{\rm Prob}}(\Delta{\ln t}<x)=\exp(-x),\label{exp_dist}
\end{equation}
where $k$ has been dropped from the notation. Finally, since 
\begin{equation}
\tau_{k}=\sum_{i=0}^{k-1}\Delta_{t_{i}}
\end{equation}
is a sum of $k$ independent exponential variables with unit average,
it follows that $\tau_{k}$ is Gamma distributed, with density given
by 
\begin{equation}
P_{\tau_{k}}(t)=\frac{t^{k-1}}{(k-1)!}e^{-t}.
\end{equation}
For large $k$, the Gamma distribution can be approximated by a Gaussian,
and the waiting time $t_{k}$ is therefore approximately log-normal.

\begin{figure}[!t]
\hfill{}\includegraphics[bb=15.3bp 183.6522bp 558.45bp 596.87bp,clip,width=1\columnwidth]{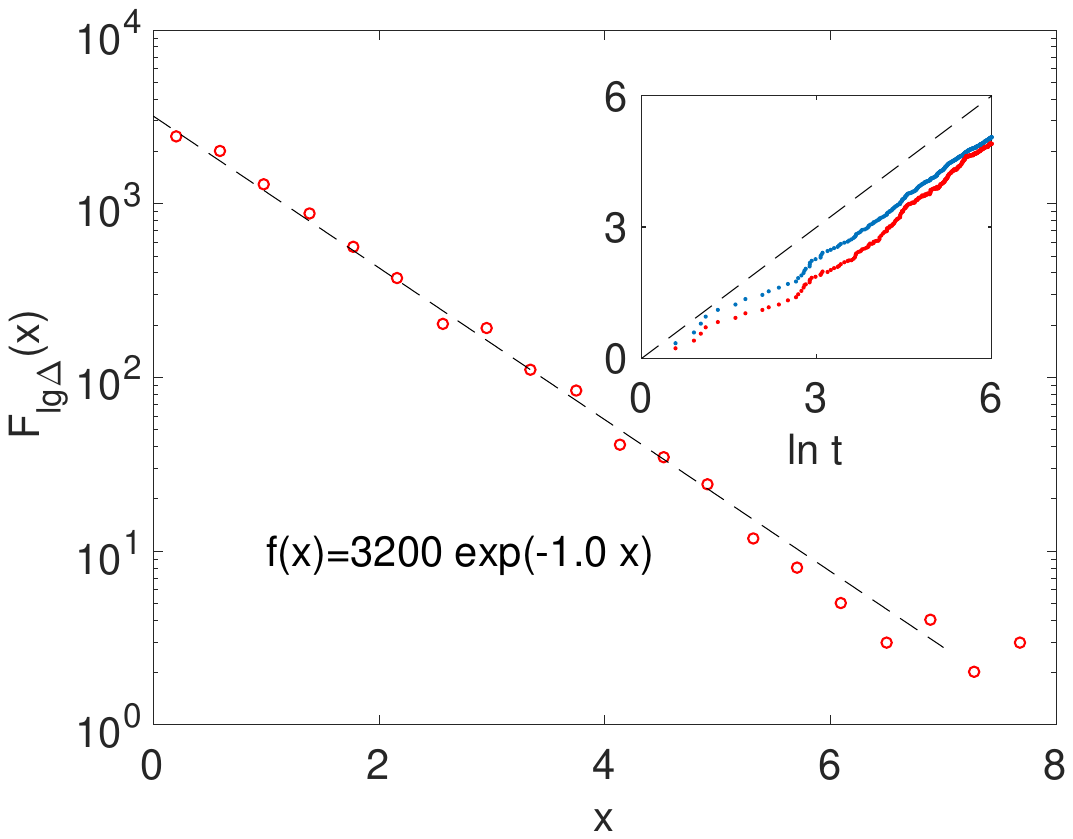}\hfill{}\caption{The circles show the unnormalized distribution of log-waiting times
in a record sub-series. The hatched line is the exponential fit given
in the figure text. The insert shows the mean and variance (blue and
red dots) of the number of records vs. the logarithm of time. The
hatched line has unit slope. { {red} The statistics describes records in 
white noise time series, constructed from independent Gaussian deviates, as explained in the main text. }}
\label{fig:Rstat}
\end{figure}

{ {red}To justify the mathematical
approximations behind Eqs.~\eqref{logPoisson}, i.e. mainly replacing 
sums  with integrals, 
we consider $500$
series each comprising $50000$ independent Gaussian deviates with
zero mean and unit variance. To create a time variable, time intervals
between the adjacent elements of each series are drawn from a uniform
distribution in the unit interval. The actual record sub-sequences
of each stationary time series are extracted and analyzed statistically.
Figure~\ref{fig:Rstat} shows that the log waiting times are exponentially
distributed, as claimed. The insert shows that the average and variance
of the number of records are nearly proportional to the logarithm
of time, in fair agreement with Eq.~\eqref{mean_var}.}

To conclude, the properties of record statistics of noisy data are
expectedly quite general, as the the noise statistics does not enter,
with exception of the independence of successive noise events.
\subsection{\;\; Quake statistics in real data}

\label{Quakes} In applications, the time series $\mathbf{S}$ will
often be a physical signal describing pseudo equilibrium fluctuations,
e.g. energy fluctuations at constant temperature. The sub-sequence
$\mathbf{R}$ is not directly available as record signal, but is extracted
from $\mathbf{S}$ by heuristically identifying the anomalous events,
or quakes, leading from one pseudo-equilibrium state to the next.
A statistical analysis is then used to ascertain if $\mathbf{R}$
behaves as the record signal of $\mathbf{S}$. 

The analysis must  consider the spatial structure from which the signal originates.
A  spatially extended system with
a finite correlation length can be treated as composed of $M$ dynamically
independent domains, with $M$ proportional to the system's size.
If quakes within each domain are a record signal, their average and
variance will both grow as $M\ln t$. Correspondingly, the logarithmic
rate of quakes will be $r_{lt}=M$. In a time series comprising all
quakes, the log-waiting times between successive events will mix inputs
from different domains and their PDF will \emph{not} be given
by $F_{\Delta{\ln t}}(x)=\exp(-Mx)$. \;\; Equation~\eqref{exp_dist}
is only valid when the log-waiting times $\ln t_{k}-\ln t_{k-1}$
are constructed from quakes occurring in the same domain. If this
spatial distinction is neglected, the time differences $t_{k}-t_{k-1}$
rather than the logarithmic time differences $\ln t_{k}/t_{k-1}$
will be exponentially distributed. This property allows one to estimate
the domain size as the largest spatial domain from whose dynamical
data comply with Eq.\eqref{exp_dist}, see e.g.~\cite{Sibani19}.

\subsection{\;\; Power laws from RD\label{subsec:-Power-laws}}

Power laws express an underlying scale invariance and are naturally
connected to critical behavior. In some cases, however, scale invariance
is not associated to length scales in real space, but to free energy
scales in configuration space. To be concrete, thermal relaxation
in a self-similar energy landscape with valleys within valleys, as
in Fig.~\ref{fig:FLandscape}, on all scales is described by power
laws~\cite{Grossmann85,Hoffmann88,Sibani89} over a range of temperatures
extending from zero up to a critical value~\footnote{Logarithmic oscillations appear when the scale invariance of the landscape
has discrete character.}.

RD is a coarse-grained description of relaxation in a hierarchy, and
produces power-laws with no connection to criticality. The details
can differ, but stem in all applications from log-time homogeneity.
Since we will show below that RD is a log-Poisson process, we consider
its effects on different observables without further ado here. The
first example, further discussed in \ref{TNM}, is the survival probability
$P(t|t_{{\rm w}})$ that a species extant at time $t_{{\rm w}}$ is
still extant at $t>t_{{\rm w}}$, where times are measured in a suitable
unit. A natural assumption is that each quake impinging on the system
reduces the survival probability by the same factor, i.e. as a function
of the number $n$ of quakes, 
\begin{equation}
P(n)=x^{n},
\end{equation}
with $0<x<1$. In order to extract a time dependence, the expression
must be averaged over the Poisson distribution of the number of quakes
falling in the observation interval $(t_{{\rm w}},t)$. With the constant  logarithmic
quaking rate coefficient denoted by $r_{{\rm q}}$, this yields 
\begin{eqnarray}
P(t|t_{{\rm w}}) & = & \left(\frac{t}{t_{{\rm w}}}\right)^{-r_{{\rm q}}}\sum_{n=0}^{\infty}\frac{(xr_{{\rm q}}\ln(t/t_{{\rm w}}))^{n}}{n!}\nonumber \\
 & = & \left(\frac{t}{t_{{\rm w}}}\right)^{-r_{{\rm q}}(1-x)}.\label{plr}
\end{eqnarray}
This is an example of pure aging and of a power-law whose exponent
$\lambda=-r_{{\rm q}}(1-x)$ is unrelated to critical behavior.

Consider now the coarse-grained autocorrelation function of, say,
the energy, between times $t_{{\rm w}}$ and $t$. Because of log-time
homogeneity, it has the generic form~\cite{Sibani05} 
\begin{equation}
C_{E}(t,t_{{\rm w}})=\sigma^{2}\sum_{k}w_{k}e^{\lambda_{k}(\ln t-\ln t_{{\rm w}})}=\sigma^{2}\sum_{k}w_{k}\left(\frac{t}{t_{{\rm w}}}\right)^{\lambda_{k}},\label{corr}
\end{equation}
where $\lambda_{k}<0$ are the eigenvalues of the (unknown) master
equation describing the process and $w_{k}>0$ are the corresponding
positive spectral weights.

\section{\;\; Applications}

\label{App}

\subsection{\;\; Punctuated Equilibrium and Ecosystem dynamics}

Gould and Eldredge first suggested~\cite{Eldredge72} and then vigorously
argued~\cite{Gould77,Eldredge88} that the abrupt changes observed
in the fossil record, i.e. mass extinctions, are a significant mode
of evolution, rather than the outcome of random fluctuations. Their
thesis, dubbed \textit{Punctuated Equilibrium}, is reconciled with
Darwinian phyletic gradualism in Stephen J. Gould's opus magnum~\cite{Gould02},
\emph{The Structure of Evolutionary Theory}.

Raup and Sepkoski~\cite{Raup82} analyzed mass extinctions in the
marine fossile record and proposed that the apparent decline in their
rate since the Cambrian explosion was the consequence of an on-going
optimization process. This idea was later taken up in~\cite{Sibani95,Sibani98a},
where extinction events were described in terms of jumps in a rugged
fitness landscape, and the decay was linked to record statistics.

Newman and Eble~\cite{Newman99c} showed that the decline in the
extinction rate during the Phanerozoic is accurately described by
a logarithmic increase of the cumulated number of extinct families
vs. time, measured since the Vendian-Cambrian boundary at about 544
Ma. This observation, we now belatedly realize, fully concurs with
an RD description of the evolution process. Newman and Eble's analysis
is discussed in more detail in the next section.

Punctuated Equilibrium has also been interpreted~\cite{Bak93,Bak97}
as a manifestation of Self-Organised Criticality, originally a stationary
paradigm for complex behavior of slowly driven systems~\cite{Bak97a}.
Along this line, Bak and Sneppen~\cite{Bak93} use a simple and elegant
model, where a set of "species" comprises an array of random numbers
between zero and one, each considered to be a fitness measure. Each
species interacts with neighbors, say two neighbors for simplicity.
The dynamics iterates two steps: a removal of the species with the
\emph{lowest} fitness and its two neighbours, followed by the replacement
of the removed, i.e. %
\mbox{%
``extinct '',%
} species with three new randomly drawn numbers. The removal of the
two neighbors represents the interdependence between co-evolving species.
Note that the number of species remains constant in time.

The model soon enters a state in which the fitness values of extant
species are mainly concentrated above a certain threshold value. Every
so often, a newly introduced species will possess a fitness below
this threshold and a burst of extinctions occurs until all fitness
values have returned above the threshold. The number of extinction
and replacement events needed to return the system above threshold
has a broad distribution and in this sense the model exhibits features
of SOC.

Whether the Bak-Sneppen model exhibits Punctuated Equilibrium or not
is in the eyes of the beholder. The extinction and renewal activity
is always occurring at the same rate, namely one, the species of lowest
fitness, plus the assumed number of its neighbours, which is the same
for all species. Bak and Sneppen, however, argued that the real. evolutionary
time scale is not given directly by the updates, but is an exponential
function of how far below the threshold a given lowest fitness value
is. As the update activity is considered on this supposedly true macroscopic
time, punctuation is introduced because of the assumed exponential
blow up. It is perhaps not entirely satisfactory that the main feature
the model is supposed to reproduce hinges on an exponential relation
to blow up the waiting time between abrupt events. Such a relationship
is seen in thermal activation over energy barriers, although being
far below the threshold should produce a faster rather than slower
return to average behavior.

In the following we discuss evolution processes with very different
time scales involving the fossil record, bacterial cultures in a flask,
a model of ecosystem evolution and the sequence of exit times of ants
from their nest. In the last case, both observational data and a model
linking them to spatial rearrangements of interacting ants are discussed.

\subsubsection{\;\; Macroevolution and the fossil record}

\label{FE} 
\begin{figure}[t]
\hfill{}\includegraphics[bb=0bp 193.3922bp 595bp 673.154bp,clip,width=0.98\columnwidth]{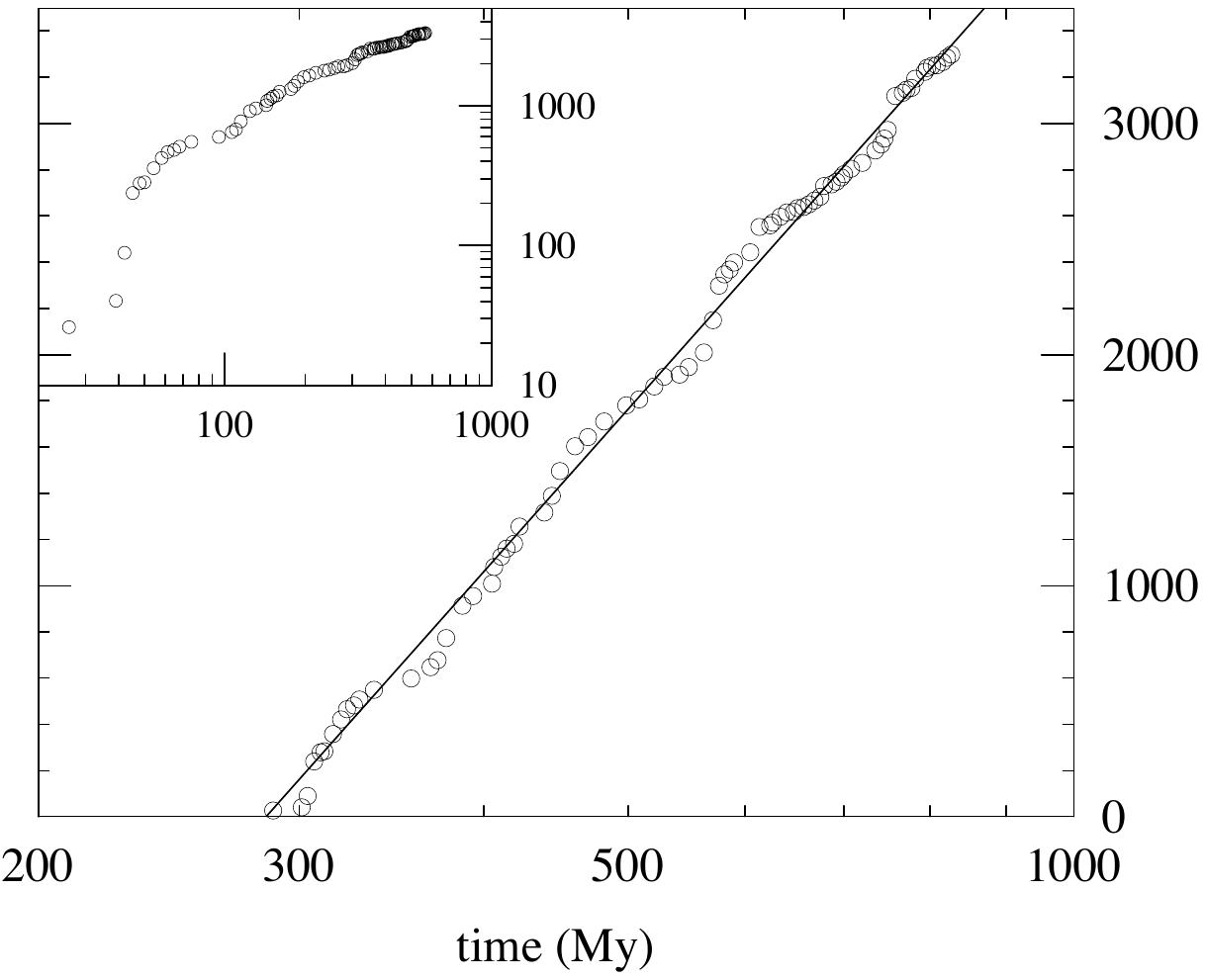}\hfill{}
\caption{Main figure: the cumulative extinction intensity as a function of
time during the Phanerozoic on linear--log scales. The straight line
is the best logarithmic fit to the data. Inset: the same data on log--log
scales. Figure taken from Newman and Eble~\cite{Newman99c}.}
\label{accum}
\end{figure}

The analysis by Newman and Eble~\cite{Newman99c} of
marine families extinction, on which this section is based and from
which Fig.~\ref{accum} is taken, argues that species extinction
is a decelerating process, and suggest a date at which this process
began, i.e. 260 My before the Cambrian explosion. We include these
important results together with equally important data from bacterial
evolution~\cite{Lenski94,Wiser_et_al_2013}, to highlight the similarity
of evolutionary dynamics on very different time scales and to suggest
an explanation in terms of RD. The original work should be consulted
for more details. 

Figure~\ref{accum} shows the cumulated number of extinctions of
marine families, as a function of time. The cumulative extinction
appears as a straight line when plotted on log-linear axes. Note that
the time axis is inverted. The fit provided is of the form 
\begin{equation}
c(t)=A+B\log(t-t_{0}),\;t>t_{0}\label{logfit}
\end{equation}
where the `initial time' s given as $t_{0}=-262$~My. The insert
shows that a log-log plot does not lend itself to a linear fit, and
that a power-law description is unsuitable.

The fossil record will never provide decades of data on a scale of
million of years. In spite of this obvious limitation, the quality
of the fit is confirmed by the presence of small logarithmic oscillations,
and the conclusion that macro evolution is not a stationary process,
a property of SOC type models~\cite{Bak93,Bak97}, seems inescapable.

Without discussing specific dynamical models, we note that extinction
events are macroscopic irreversible changes whose cumulated number
grows logarithmically in time. Both these properties characterize
quakes in RD and are found in the Tangled Nature Model described below.

\subsubsection{\;\; Bacterial populations in a controlled environment}

\label{BIF} 
\begin{figure}[t]
\hfill{}\includegraphics[clip,width=0.98\linewidth]{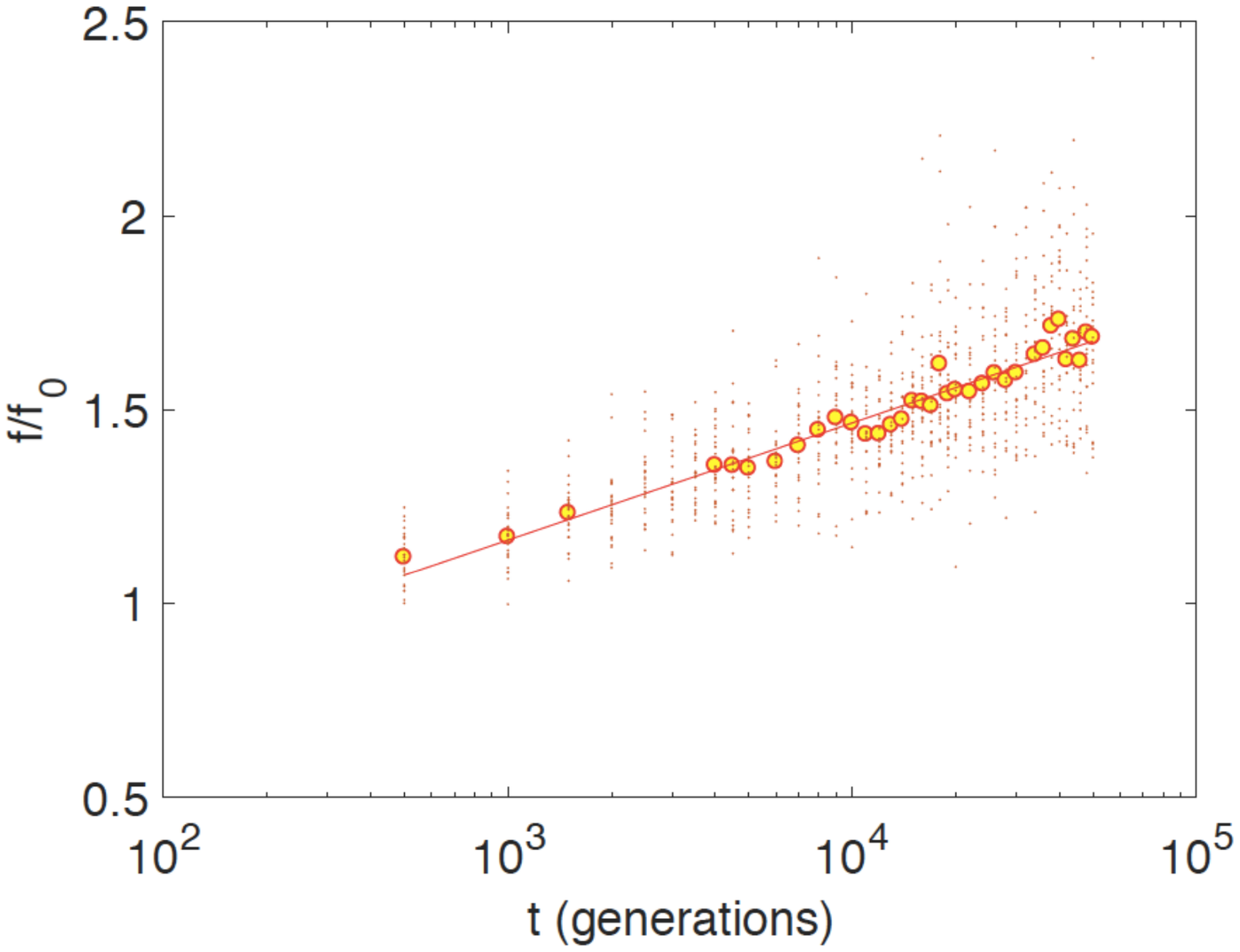}\hfill{}
\caption{The points depict the Malthusian fitness of 12 E. coli monoculture
populations, after an initial change of food substrate, as a function
of logarithmic time, data from Ref.~\cite{Wiser_et_al_2013}. Circles
are the average fitness of the populations.}
\label{Lenski-Fig}
\end{figure}

Lenski and Travisano~\cite{Lenski94} and later Weiser
et al.,~\cite{Wiser_et_al_2013} measured the Malthusian fitness
of 12 E. coli populations grown in a constant physical environment
over 10000 and 50,000 generations, respectively. A generation begins
when a bacterial sample is injected in fresh substrate and ends when
its population reaches its first plateau. In the experiments, 12 initially
identical cultures follow different evolutionary paths, all leading
to increasing cell size and increasing Malthusian fitness, i.e. initial
rate of growth of the population. In the lack of competition with
other organisms, all evolutionary changes must concern the metabolic
pathways needed to process the substrate. The situation can be modelled
by an adaptive walk on a rugged fitness landscape~\cite{Kauffman87}
where any improvement, i.e. a fitness record, quickly moves the population
to a higher fitness peak~\cite{Sibani99a}, i.e. induces the macroscopic
change we presently call a quake. Assuming that each quake leads to
the same improvement, RD predicts a logarithmic Malthusian fitness
increase of the bacterial populations. Figure~\ref{Lenski-Fig} depicts
the Malthusian fitness of the 12 populations as dots and their average
as circles. The logarithmic trend of the data is clear. It is equally
clear that a logarithmic trend cannot continue indefinitely and that
the growth law $f=f_{0}(1-\exp(-\alpha t))$, where the exponent $\alpha$
is a positive number close to zero, grows as a logarithm for $t\ll1/\alpha$
and remains bounded for $t\rightarrow\infty$. 

In summary, the Malthusian fitness of E. choli evolving in a fixed
environment undergoes a decelerating process similar in certain respects
to evolution on much larges scales. The situation is well described
by the Kauffman model~\cite{Kauffman87}, where the fitness of an
individual is fully determined by its genome. This is the opposite
of the Tangled Nature Model, described below, where fitness is determined
by a tangle of interactions connecting individuals of different species.
Interestingly, both models feature logarithmic growth laws.

\subsubsection{\;\; A model of ecosystem dynamics}

\label{TNM} The agent based Tangled Nature Model~\cite{Christensen02,Collabiano03}
captures key elements of ecosystem evolution, including punctuated
equilibrium~\cite{Christensen02} and a power-law distributed life-time
of species~\cite{Andersen16}. Over the last two decades a whole
family of Tangled Nature models appeared, a development recently described
in~\cite{Jensen18}, where applications to different fields of science
are reviewed.

Closely related to our current RD focus, the model features qualitative
structural changes or quakes which can be interpreted as extinctions.
To a good approximation these are uniformly distributed on a logarithmic
time axis~\cite{Becker14}, i.e. the logarithmic rate of quakes is
constant, as predicted by RD. Correspondingly, the integrated number
of quakes grows with the logarithm of time, which concurs with the
behavior extracted from the fossil record~\cite{Newman99c}. Punctuations,
i.e. \emph{quakes}, irreversibly disrupt \emph{quasi-Evolutionary
Stable Strategies} (qESS), periods of metastability where population
and the number of extant species, or diversity, fluctuate reversibly
around fixed values.

The basic model variables are binary strings of length $L$, i.e.
points of the $L$ dimensional hypercube. Variously called \emph{species}
or \emph{sites}, these are populated by \emph{agents} or \emph{individuals},
which reproduce asexually in a way occasionally affected by random
mutations. Only a tiny fraction of the possible sites ever becomes
populated during simulations lasting up to one million generations.
The \emph{extant species}, i.e. those with non-zero populations at
a given time, are collectively referred to as \emph{ecosystem}, and
their number as \emph{diversity}. Unlike e.g. the Kauffman model~\cite{Kauffman87},
an agent's reproductive success is not predicated on its genome, but
the interactions, or couplings, it has with other agents.

How agent $b$ affects the reproductive ability of agent $a$ and
vice-versa only depends on the species to which these agents belong.
The interaction is non-zero with probability $\theta$, in which case
the couplings ($J_{ab}$, $J_{ba}$) are extracted, once and for all,
from a distribution well described by the Laplace double exponential
density $p(x)=\frac{1}{2a}e^{-\lvert x-\overline{x}\rvert/a}.$ For
the data shown, the parameters $\overline{x}$ and $a$ are estimated
to $-0.0019$ and $0.0111$, respectively. Note that the long term
dynamics of the model strongly depends on the coupling distribution
having finite or infinite support See Refs.~\cite{Becker14,Andersen16}
for further details.

The set of interactions linking an individual to others is key to
its reproductive success and arguably constitutes its most important
property. Yet, in many studies, e.g.~\cite{Anderson05,Jones10,Becker14},
the interactions of an individual and those of its mutated off-spring
are unrelated, a rather unrealistic feature corresponding to a point
mutation turning a giraffe into an elephant.

This issue has been addressed~\cite{Sevim05,Laird06} by introducing
correlated interactions between parent and off-spring. More recently,
a family of models was introduced~\cite{Andersen16} , parametrized
by a positive integer $K$, where $K=1$ is the original version,
with uncorrelated interactions, and where the degree of correlation
grows with $K$. The main effect of correlated interaction is seen
in the species survival probability, described in Fig.\ref{fig:t-on-tw}.

Let $\mathcal{S}$, $N_{b}(t)$ and $N$ denote the ecosystem, the
population size of species $b$, and the total population {$N(t)=\sum_{b}N_{b}(t)$}.
An individual of type $a$ is chosen for reproduction with probability
$n_{a}=N_{a}/N$, and succeeds with probability $p_{{\rm off}}(a)=1/(1+e^{-H_{a}})$,
where 
\begin{equation}
H_{a}(t)=-\mu N(t)+\sum_{b}j_{ab}(t),\label{eq:Pfunc}
\end{equation}
and where 
\begin{equation}
j_{ab}=\frac{N_{b}}{N}J_{ab}=J_{ab}n_{b}\label{dens_w_c}
\end{equation}
is a density weighted coupling. In Eq.~\eqref{eq:Pfunc}, $\mu$
is a positive constant which represents the finiteness of resource
availability curbing population growth. Letting $p_{{\rm mut}}$ be
the mutation probability per bit, parent and offspring differ by $k$
bits with probability Bin$(k;K,p_{{\rm mut}})$, the binomial distribution.
Death occurs with probability $p_{{\rm kill}}$ and time is given
in \emph{generations}, each equal to the number of updates needed
for all extant individuals to die. Thus, with population $N$ at the
end of the preceding generation, the upcoming generation comprises
$Np_{{\rm kill}}$ updates. The parameters used are always $L=20$,
$\mu=0.10$, $\theta=0.25$, $p_{{\rm kill}}=0.20$, $p_{{\rm mut}}=0.01$,
and the initial condition invariably consists of a single species
populated with 500 individuals.

\begin{figure}[htb]
\hfill{}\includegraphics[bb=7.4375bp 171.0777bp 580.125bp 669.435bp,clip,width=0.45\textwidth]{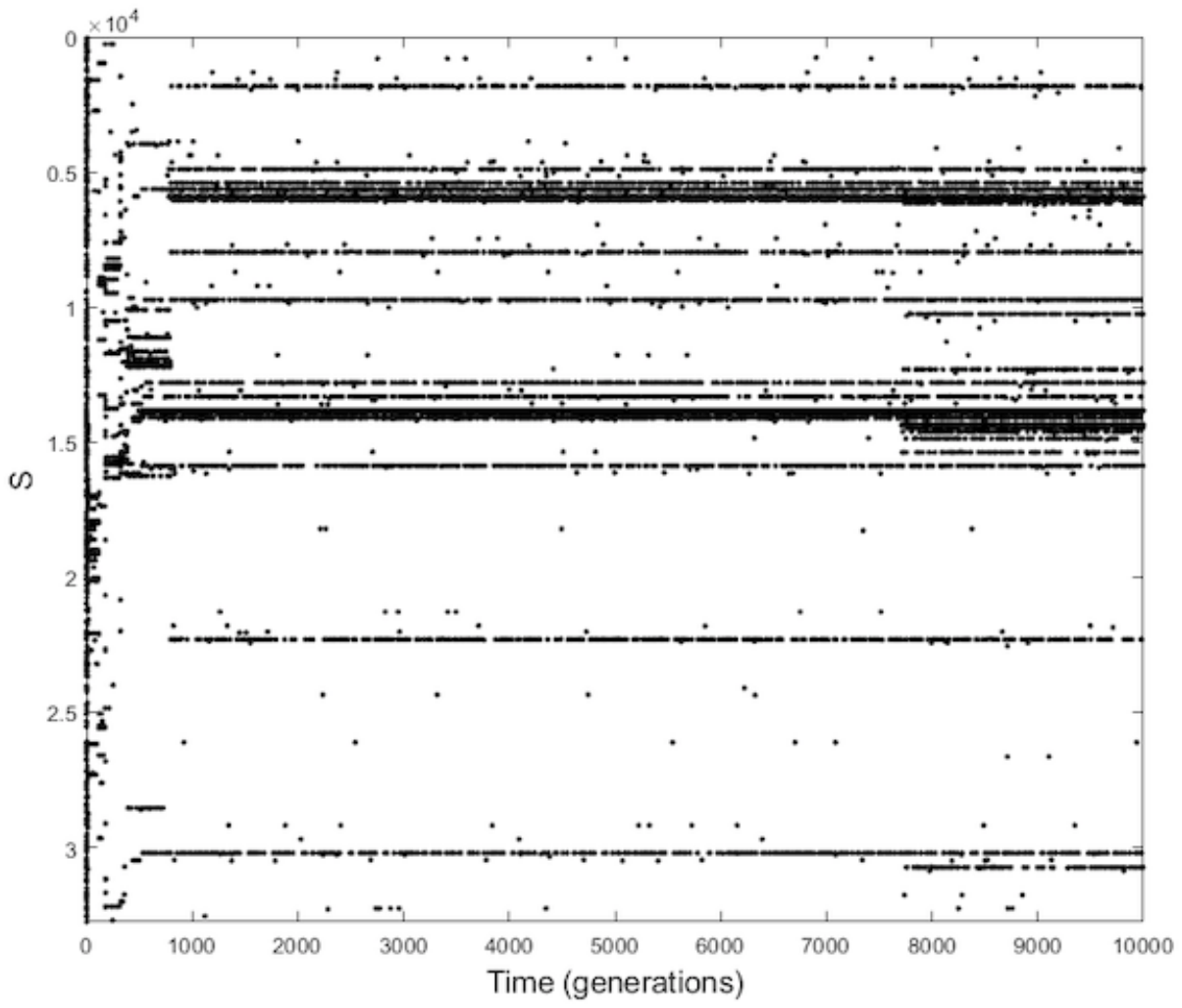}\hfill{}
\caption{A qualitative view of the evolution of a system ecology. Each extant
species is represented by a black point at the integer value corresponding
to its binary string. The abscissa is time. Long periods of stability
are punctuated by rapid changes of the extant population. Figure taken
from~\cite{Jensen18}.}
\label{fig:intermittent}
\end{figure}

The concepts of \emph{core} and \emph{cloud} species are key to define
quakes and understand the entropic mechanism causing the increasing
duration of the qESS. A core is a group of species linked by mutualistic
interactions. As a consequence, the majority of the agents resides
in the core. The cloud consists of many species, each having few individuals,
often mutants of core species.

A quake is an event which rapidly changes the composition of the core,
e.g. by replacing some or all its component species. The event is
initiated by a new mutant cloud species which receives strong positive
interactions from the core. As a consequence, it enjoys great reproductive
success and its growing population destabilizes the core via the global
negative feed-back associated to the $\mu$ term of Eq.\eqref{eq:Pfunc}.

The growing duration of qESS reflects an entrenchment into metastable
configuration space components of increasing entropy: It becomes increasingly
difficult to generate by mutations a new cloud species able to threaten
the core. The step requires an increasing number of mutations as the
system ages.

\begin{figure}[htb]
\hfill{}\includegraphics[width=1\columnwidth]{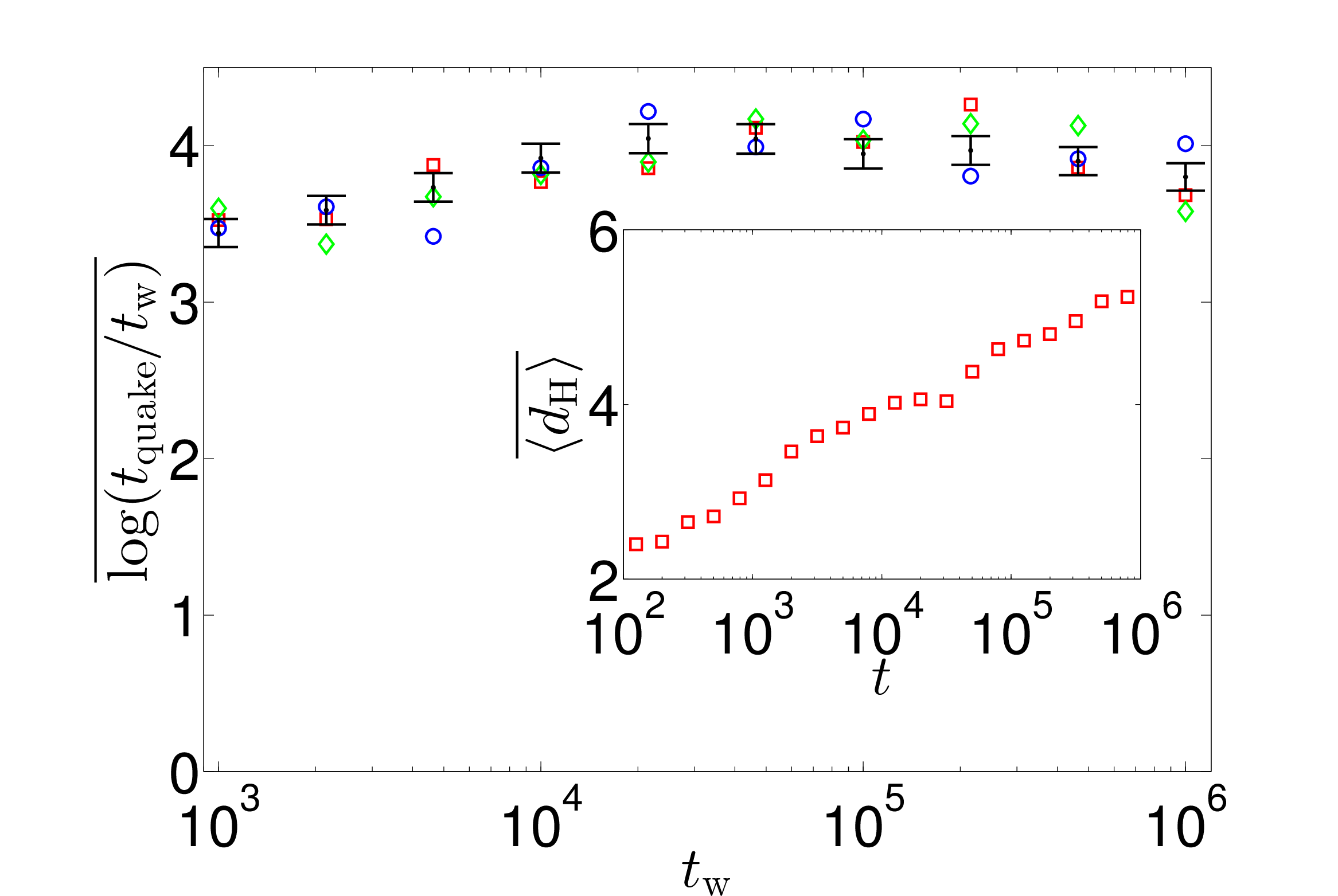}\hfill{}
\caption{Main plot: The average logarithmic waiting time $\log(t_{{\rm quake}}/t_{{\rm w}})$
is estimated using $2022$ trajectories and plotted (black) vs. $t_{{\rm w}}$,
with $1\sigma$ statistical error bars. The blue circles, green diamonds
and red squares are based on different sub-samplings and illustrate
the statistical variation of the data. Insert: Hamming distance from
cloud species to the most populous core species, plotted on a log
scale and averaged twice: over the cloud species and over $2022$
trajectories.  Figure taken from \cite{Becker14}.}
\label{fig:cloud_size_av} \vspace{-0.3cm}
\end{figure}

\begin{figure}
\hfill{}\includegraphics[width=1\columnwidth]{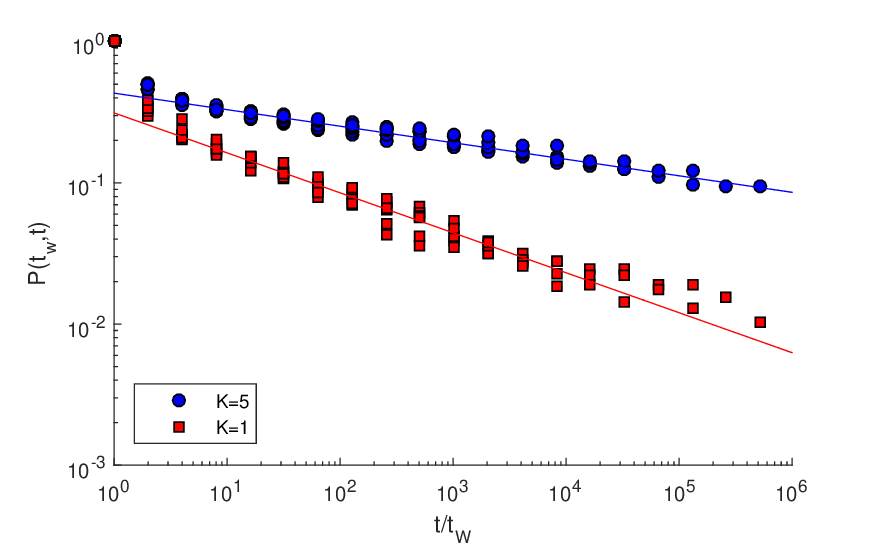}\hfill{}\caption{(Color on line) Species persistence data plotted as a function of
$t/t_{w}$. $K=1$ data are shown by squares, and $K=5$ data by circles.
The lines are least squares fits to power-laws $y=a(t/t_{w})^{b}$.
where $b=-0.283(14)$ and $b=-0.117(6)$ for $K=1$ and $K=5$, respectively.
All available cohorts have been used in the fits. Figure taken from
Ref.~\cite{Andersen16}}
\label{fig:t-on-tw}
\end{figure}

In the main panel of Fig.~\ref{fig:cloud_size_av}, the logarithmic
waiting times $\log(t_{{\rm quake}}/t_{{\rm w}})$ for ouakes falling
after time $t_{{\rm w}}$ , averaged over an ensemble of $2022$ repetitions,
are plotted vs. $t_{{\rm w}}$ with $1\sigma$ error bars. These average
log-waiting times are seen to be nearly independent of $t_{{\rm w}}$,
which indicates that quake dynamics is log-time homogeneous, as RD
posits. The insert shows the average Hamming distance between cloud
species and the most populous core species. The distance grows as
the logarithm of time, which implies that core mutants become on average
gradually less fit, and . more mutations are needed to generate a
viable mutant which can destabilize the core. Agents die at a constant
rate, and have a finite expected life-time. In contrast, as we shall
see, species go extinct at a decelerating rate, and their life-times
do not possess finite averages.

A cohort is defined as the set of species extant at time $t_{w}$
and their persistence $P(t_{w},t)$ as the fraction of the cohort
still extant at time $t>t_{w}$. Persistence provides an estimate
of the probability that a species extant at time $t_{w}$ still is
extant at later times. The distinction between the two quantities
is glossed over in the following. The life-time probability density
function of a species extant at $t_{w}$ is then 
\begin{equation}
S(t_{w},\tau)=-\frac{d}{d\tau}P(t_{w},t_{w}+\tau)\quad0\le\tau<\infty.\label{survival}
\end{equation}

Figure~\ref{fig:t-on-tw}, taken from Ref.~\cite{Andersen16} shows
that persistence data obtained for different $t_{{\rm w}}$ collapse
when plotted as function of $t/t_{{\rm w}}$, a so called pure aging
behavior. The two data sets describe models with uncorrelated and
correlated interactions between parent and off-spring ($K=1$ and
$K=5$, respectively.) As expected, inheritance of interactions leads
to higher persistence.

The lines are fits to a power-law $y(t/t_{w})=a(t/t_{w})^{b}$, For
$K=1$ the exponent is $b=-0.283(14)$ and for $K=5$ it is $b=-0.117(6)$.
The behavior is as predicted by RD, see e.g. Eq.\eqref{plr} Three
comments are in order: first, independently of the degree of inheritance,
Eq.~\eqref{survival} shows that the life-time distribution lacks
a finite average. Second, we see that species created at a late stage
of the evolution process (large $t_{w}$) are more resilient than
those created early on, implying that the rate of quakes decreases
in time. Third, the exponent of the persistence decay is more than
halved when $K$ goes from $1$ to $5$, clearly showing that inheritance
produces a more robust ecology where species live longer.

\begin{center}
\begin{figure*}[t]
\hfill{}\includegraphics[bb=0bp 105.3097bp 842bp 500.221bp,clip,width=0.8\textwidth]{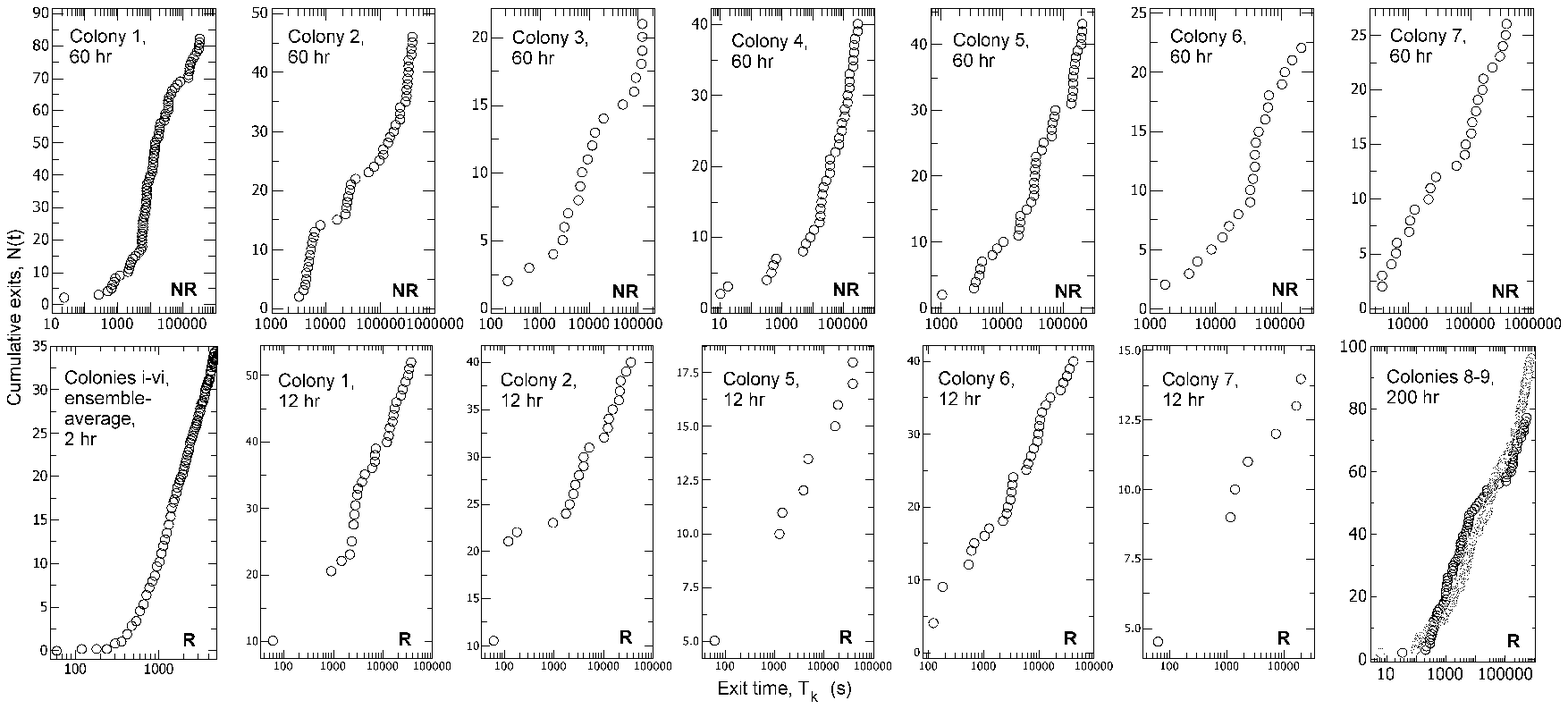}\hfill{}
\caption{The panels show the number of nest exits prior to time $t$ under
different conditions versus $t$ on a logarithmic abscissa. removal
(circles) and non-removal. Figure taken from Ref.~\cite{Richardson10}.}
\label{fig:RichAnt}
\end{figure*}
\par\end{center}

\subsubsection{\;\; Movement in ant colonies}

That biological systems comprising many interacting agents evolve
through punctuations is not narrowly dependent on the type of interaction.
Social interactions in ant colonies of \emph{Temnothorax albipennis}
seem to generate the non-stationary ant motion behavior studied by
Richardson et al.~\cite{Richardson10} and described by an RD analysis
of statistical data. Our fig.~\ref{fig:RichAnt} is taken from this
reference, which should be consulted for further details. Importantly,
the analysis of~\cite{Richardson10} is based on observational data
and hence independent of modeling assumptions. A theoretical model
of the same phenomenology~\cite{Sibani11a} is also discussed below.

In the experiment, fifteen \emph{T. albipennis} colonies were housed
in cardboard nests kept at a constant temperature ($24^{\circ}$C),
with continuous lighting. All colonies hosted a queen and had a complement
of brood at various stages of development. An ant's first exit from
the nest was recorded as an event and to ensure that only first exits
were counted the ant was either removed or allowed to return after
all ants were initially tagged and the tag detected upon exit. The
PDF of the logarithmic waiting time between successive exits is exponential,
as expected in RD and the cumulated number of exits grows proportionate
to the logarithm of time. This behavior is shown in Fig.~\ref{fig:RichAnt}
for a number of different measurements. In the lower right corner
of each panel, NR stands for "Non Removal and R for "Removal",
which refers to the fate of the ants exiting the nest.

All in all, a RD description of ant movement in \emph{T. albipennis}
seems reasonably accurate. This conclusion was challenged by Nouvellet
et al.~\cite{Nouvellet10}, who performed new experiments where the
exit process could more simply be described by standard Poisson statistics,
the result expected if the ants move randomly and independently in
and out of their nest. The question is of course whether the two sets
of experiments really were equivalent~\cite{Richardson11}. More
important from our vantage point is the mechanism producing the gradual
\emph{entrenchment} of dynamical trajectories in more long-lived metastable
configurations.

\begin{figure}
\hfill{}\includegraphics[width=0.95\columnwidth]{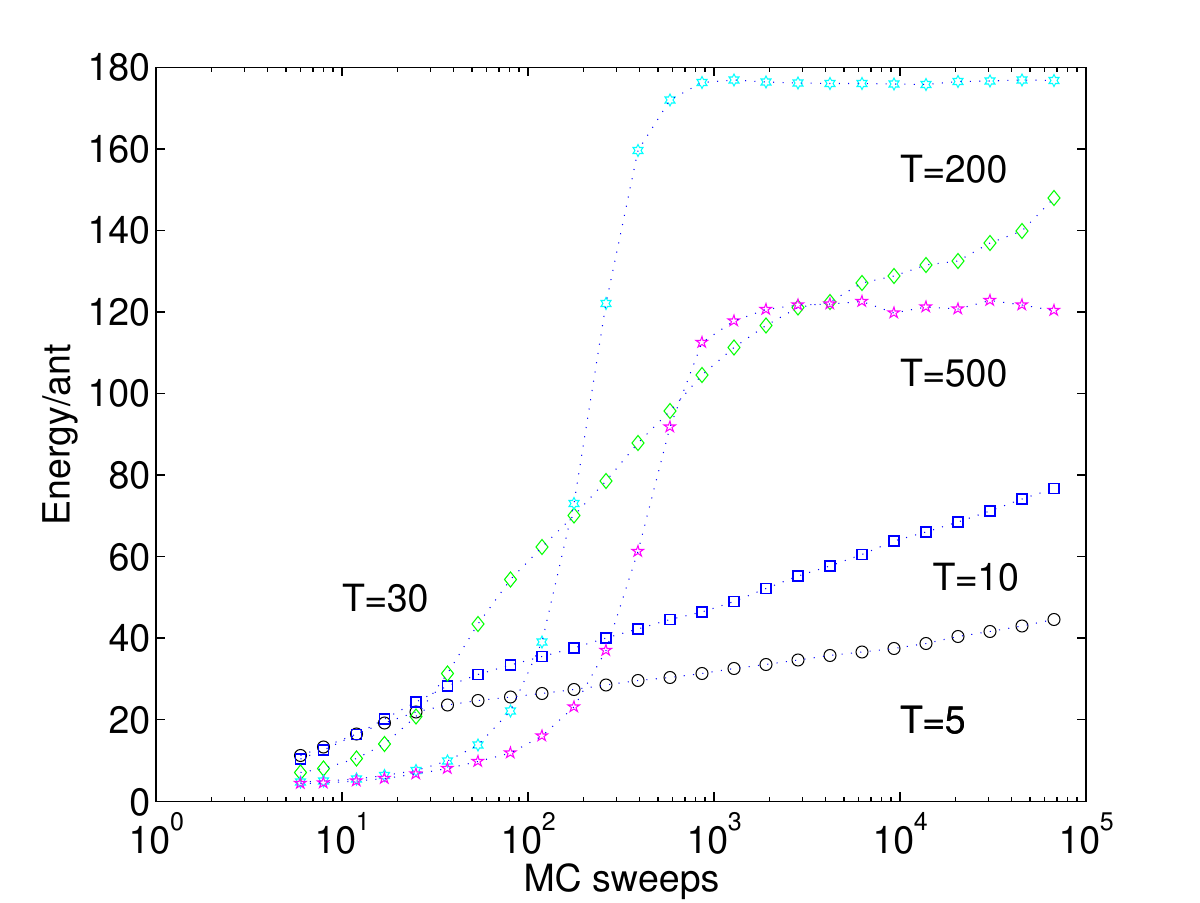}\hfill{}\caption{(Color online) All data shown pertain to a system with four ant types,
with one negative and five positive interactions between different
types. The value of the utility function per ant, averaged over $100$
trajectories, is plotted versus time. The initial configuration is
in all cases obtained by randomly placing the ants on the grid. Figure
taken from~\cite{Sibani11a}.}
\label{big_energy_plot}
\end{figure}

\begin{figure}
\hfill{}\includegraphics[bb=0bp 0bp 547.2bp 417.176bp,clip,width=0.48\columnwidth]{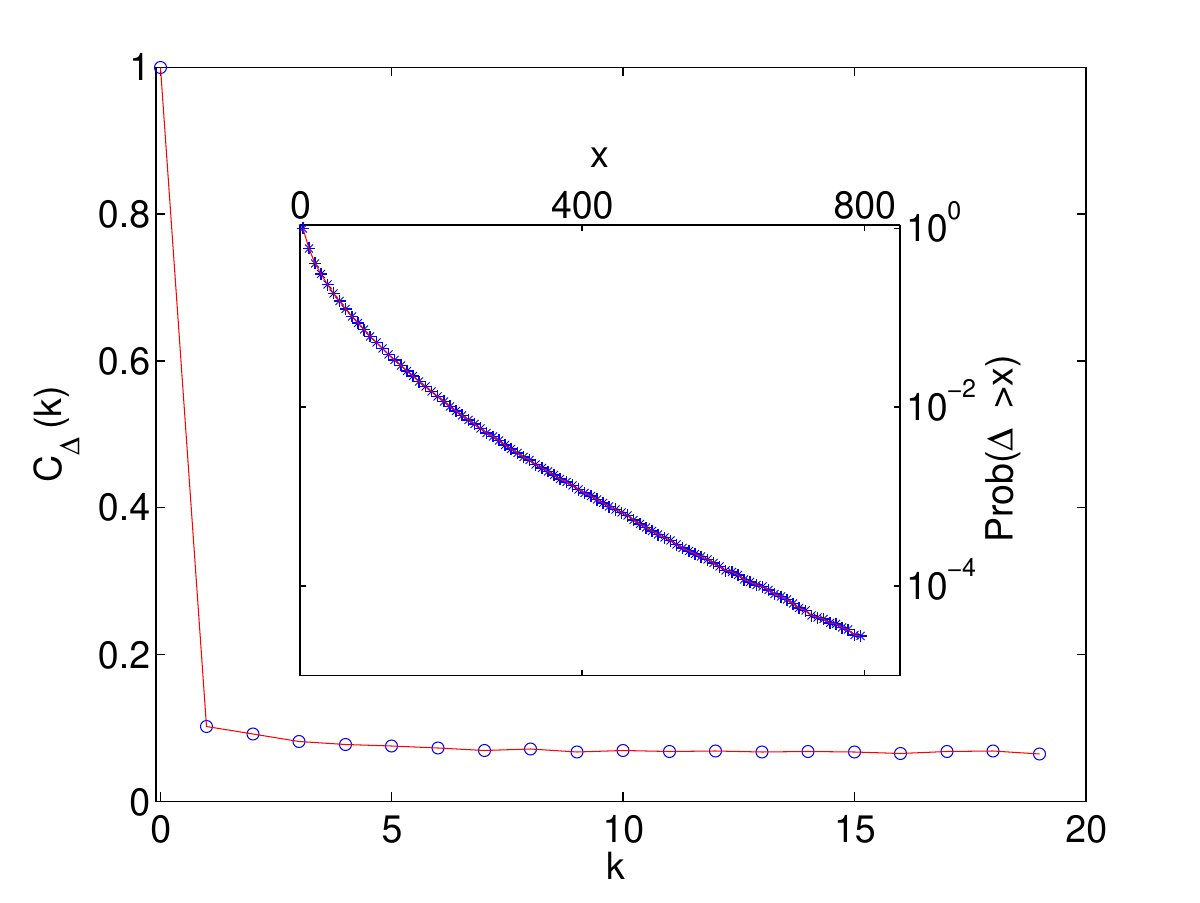}\hfill{}\includegraphics[bb=0bp 0bp 547.2bp 417.176bp,clip,width=0.48\columnwidth]{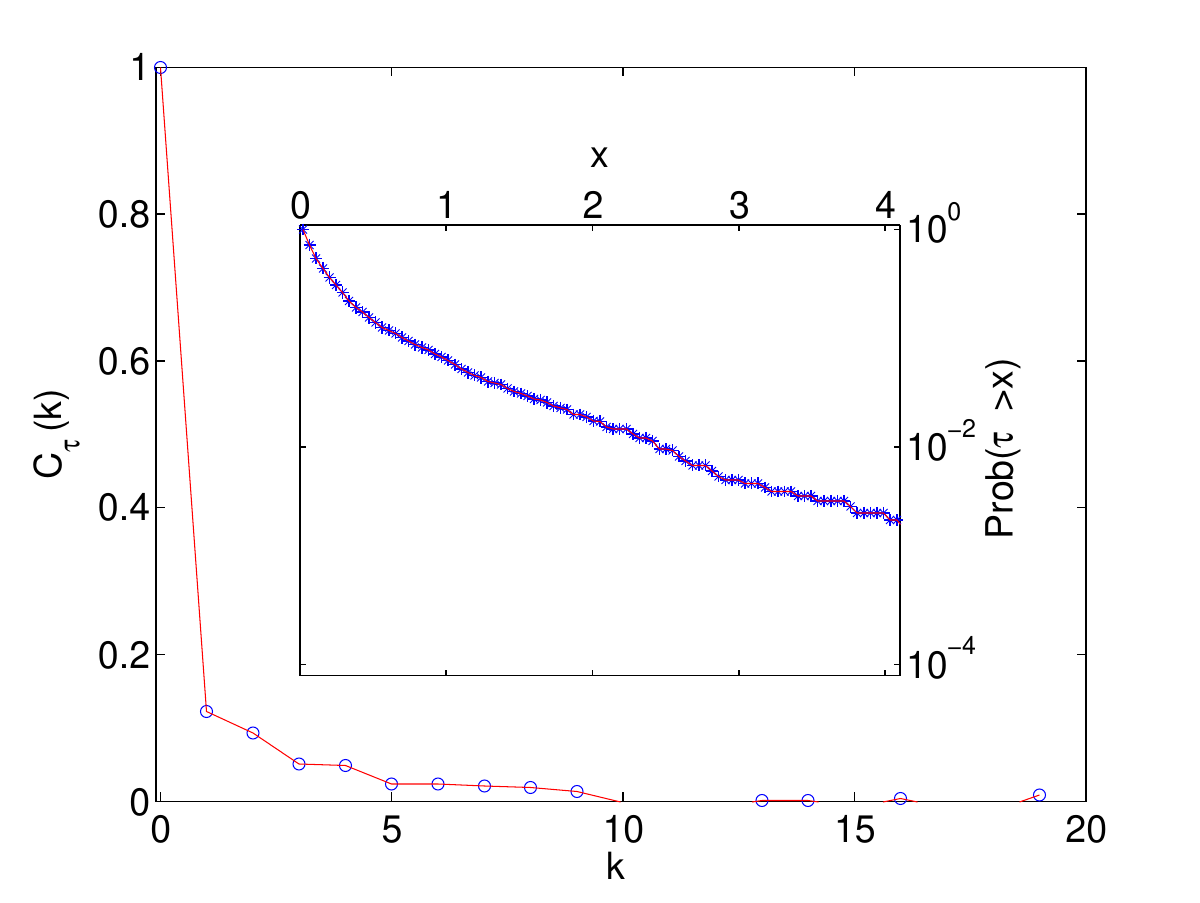}\hfill{}\caption{Left: The correlation function $C_{\Delta}(k)$ for the waiting times
between consecutive exit events, plotted vs. $k$. The insert shows
the cumulative distribution of the waiting times, plotted on a logarithmic
vertical scale. The system contains two types of ants moving on a
grid of linear size $7$ with DS parameter is $T=50$. Right: same
as above, except that the DS value is here $T=5$ and that the correlation
and cumulative distribution are calculated using the log-waiting times
rather than the waiting times. Figure taken from~\cite{Sibani11a}.}
\label{corr_and_logP}
\end{figure}

In the model proposed in~\cite{Sibani11a}, from which Figs.\ref{big_energy_plot}
and \ref{corr_and_logP} are taken, ants of different types move in
a stochastic fashion from one site to a neighboring site on a finite
2D lattice, similar to an ant's nest. Each site corresponds to a small
area which can be occupied by several ants, and one site is designed
as exit. The probability of each move is determined by the changes
of a utility function $E$ it entails. $E$ is a sum of pairwise interactions
between ants, weighted by distance. Depending on type, ants can have
positive or negative interactions, meaning that moving closer induces
a positive or negative change $\delta E$ in the utility function.
A standard Metropolis update algorithm is used~\cite{Metropolis53}
with the probability of carrying out a move given by $\min[\exp((\delta E/T),1]$,
where the parameter $T$, called `degree of stochasticity' (DS) controls
the importance of the interactions, similarly to the temperature in
a physical model. The sign inversion relative to the usual Metropolis
convention implies that movements \emph{increasing} the utility function
are unconditionally accepted.

Since interactions are negligible in the limit $T\rightarrow\infty$,
the ants act independently and their movement out of the nest are
a Poisson process in linear time, as observed by Nouvellet et al.~\cite{Nouvellet10}.
At low $T$, the systems enters a non-stationary regime, where exits
can be described as suggested by Richardson et al.~\cite{Richardson10}.
In model simulations starting from a random distribution of ants in
the nest, ever larger ant clusters establish themselves on gradually
fewer sites. This in turn creates continually growing dynamical barriers
(e.g. empty sites) for ants which have not yet joined a cluster. Whether
a similar mechanism involving social structures is at play in real
ant nests remains of course to be seen. Let us finally note that the
size of the barrier scaled during ant movement was not investigated
in~\cite{Sibani11a}. We are therefore unable to give a precise definition
of quake. and use nest exits as a proxy.

All simulational data presented in Fig.~\ref{big_energy_plot} pertain
to a system with four types of ants. The value of the utility function
per ant, averaged over $100$ trajectories is plotted vs time (i.e.
number of MC sweeps) on a logarithmic scale. For $T=500$ and $200$,
a constant value is approached. The latter increases as the DS decreases,
since the system's ability to maximize its utility function initially
improves when lowering $T$. However, for $T=5$ and $T=10$ the average
$E$ only grows logarithmically without approaching equilibrium. In
the two low $T$ curves, the mean value of the utility function is
seen to increase with $T$ rather than decreasing as in the stationary
regime. This highlights the strong non-equilibrium nature of the relaxation
process.

Figure \ref{corr_and_logP} describes the exit statistics, (corresponding
to quake statistics) in a system with two ant types. At the $k$'
th sweep the program checks whether motion has occurred at the site
dubbed `exit' and, if so, registers the corresponding time $t_{k}$.
The simulations, each running from $t=5$ to $t=5\cdot(1+10^{5})$,
are repeated $100$ times in order to improve the statistics. The
DS values used in the simulation are $T=50$ and $T=5$ in the left
and right panel, respectively. The data shown are statistically very
different in spite of being graphically rather similar. In the left
panel the waiting times, i.e. the time differences $\Delta_{k}=t_{k}-t_{k-1}$
are analyzed with respect to their correlation and their distribution.
Since the $t_{k}$' s are integer rather than real numbers, the exit
process can never be truly Poissonian. We nevertheless estimate the
normalized correlation function of the $\Delta_{k}$'s, averaged over
$100$ independent runs. For independent entries, the latter would
equal the Kronecker's delta $C_{\Delta}(k)=\delta_{k,0}$. We furthermore
estimate the probability that $\Delta>x$, as a function of $x$.
For a Poisson process this probability decays exponentially in $x$.
The correlation and probability distribution are plotted in the main
figure and the insert, using a linear and a logarithmic ordinate,
respectively. We see that short waiting times, i.e. $\Delta_{k}$'s
of order one are over-represented relative to the straight line representing
the Poisson case. Secondly, the correlation decays to about $1/10$
in a single step, but then lingers at that value. Taken together,
these two feature indicate that a short waiting time is more likely
followed by another short waiting time, i.e. that the motion often
stretches over several sweeps.

The right panel of the figure shows data obtained as just discussed,
except that logarithmic time differences $\tau_{k}=\ln(t_{k})-\ln(t_{k-1})$
rather than linear ones are utilized. The correlation function $C_{\tau}(k)$
decays quickly to zero, albeit not in a single step, and the probability
that $\tau>x$ is nearly exponential. Again, short log-waiting times
are over-represented in the distribution, and since the correlation
decays to near zero in $k=5$, they are likely to follow each other.
Thus, also in this case ant at the 'exit' stretches beyond a single
sweep. In summary, banning the effect of our time unit, the sweep,
being too short relative to the de-correlation time of ant motion,,
$T=50$ data are, as expected in a stationary regime, well described
by a Poisson process, while $T=5$ data are well described by a log-Poisson
process.

To conclude, the experiments of Richardson et al.~\cite{Richardson10}
showed that the nest exit statistics of \emph{Temnothorax albipennis}
are well described by RD. This behavior requires a hierarchy of dynamical
barriers, expectedly due to ant interactions whose details are unknown.
A simple agent based model of interacting ants has a `glassy' phase
where the interaction between the agents are important and the experimental
findings are reproduced. When the agents' motion becomes sufficiently
random, the system equilibrates, and the RD behavior disappears. The
model hints at a possible origin of the barrier hierarchy required
by RD, but the issue needs further work for clarification, as does
the biological relevance and possible mechanism controlling the degree
of stochasticity of real ants.

\subsection{\;\; High $T_{c}$ superconductivity. The ROM model}

This section deals with the motion of magnetic flux lines inside type
II superconductors. We mainly follow Ref.~\cite{Oliveira05}, which
describes simulations of the Restricted Occupancy Model (ROM). This
reference, which should be consulted for further details, presents
an early application of RD ideas and uses a notation which differs
from ours in the rest of this paper. In particular, $t$ is not the
time elapsed since an initial quench, but an observation time. For
the reader's convenience we follow in this section the notation of
Ref.~\cite{Oliveira05}. Type II superconductors are layered materials
which remain superconducting
even after an applied magnetic field $H$ begins to enter the sample
at a field strength $H>H_{c_{1}}$. When $H$ exceeds a second value
$H_{c2}>H_{c_{1}}$ the material becomes a normal conductor. In the
range $H_{c_{1}}<H<H_{c_{2}}$, the external magnetic field penetrates
into the bulk of the superconductor by forming vortex or flux lines
created by circulating currents, or vortices, of superconducting electrons.
Parallel flux lines have a repulsive interaction energy that depends
on their distance, and pushes them towards the bulk of the material.
Inhomogeneities tend to trap the flux lines and impede the rapid establishment
of a uniform flux density. A mechanical force balance is established
between the pinning force exerted by an inhomogeneity, also called
a pinning center, and the force produced by the gradient in the surrounding
flux line density. A pinning center corresponds to a local potential
energy well from which thermal fluctuations can release a flux line
and allow it to move towards regions of lower flux density, kicking
as a result other flux lines out of their traps. In this way, flux
avalanches lower the gradient of flux line density as flux lines move
into the bulk region of the material. As a consequence, the difference
between the internal magnetic field and the external applied field
decreases in a relaxation process towards thermodynamic equilibrium.

{ {red}
The ROM model introduces  a discrete grid  for each superconducting layer and  uses the number $n_i$ of 
flux lines   transversing the $i$'th plaquette of the grid  as  dynamical variables.
 The only interactions
 included  are those  between flux lines  in the same or nearest neighbour plaquettes,
 and  the motion of the flux lines is represented by a time variation of their numbers in each plaquettes.
 The energy of a flux line
  configuration $\{n_i\}$ is given by 
\begin{equation}
E =\sum_{ij}A_{ij}n_{i}n_{j}- \sum_{i}A_{ii}n_{i}+\sum_{i}A_{i}^{p} n_{i} +
\sum_{\left\langle ij\right\rangle _{z}}A_{2}\left(  n_{i}-n_{j}\right)  ^{2},
\label{ROM_hamilton}
\end{equation}
The first  term represents the repulsion energy due to vortex-vortex interaction within a  layer, and the second 
the  vortex self energy. 
As already mentioned, interactions 
beyond nearest neighbors are neglected.
 We set $A_{0}\stackrel{\rm def}{=}  A_{ii}=1$, $ A_{1}\stackrel{\rm def}{=}A_{ij}$ if $i$ and $j$ are nearest neighbors on the same layer, and $A_{ij}=0$ otherwise.
The third term represents the interaction of the vortex pancakes with the pinning centers. $A^{p}$ is a random potential and we assume for simplicity that $A^{p}$ has the following distribution $P\left(  A^{p}\right) =\left(  1-p\right)  \delta\left(  A^{p}\right)  -p\delta\left(  A^{p}-A_{0}^{p}\right)  $
The pinning strength $\left|  A_{0}^{p}\right|$ represents the total action of the pinning centers located on a site. 
In the simulations described here $ A_{0}^{p} =-0.3$.
Finally the last term in Eq. \eqref{ROM_hamilton}
describes the interactions between the vortex sections in different layers. 
This term is a nearest neighbor quadratic interaction along the $z$ axis, so that the number of vortices in neighboring cells 
along the $z$ direction tends to be the same.

The parameters of the model are defined in units of $A_{0}$. The time is measured in units of  Monte Carlo (MC) sweeps.  
We assume the external magnetic field to be applied perpendicularly    to the planes and for this reason we only consider motion 
of the vortex pancakes within the planes and use periodic boundary conditions in the $z$-direction. 
Each individual Monte Carlo update consists in selecting a vortex pancake at random 
and moving it to a randomly selected neighbour position. As always in Metropolis importance sampling the movement of the vortex
 is automatically accepted if the energy of the system decreases or remain unchanged. 
 If the energy of the system increases, the movement is accepted with probability $\exp(-\Delta E/T)$. 
}

\begin{figure}
\hfill{}\includegraphics[clip,width=8cm]{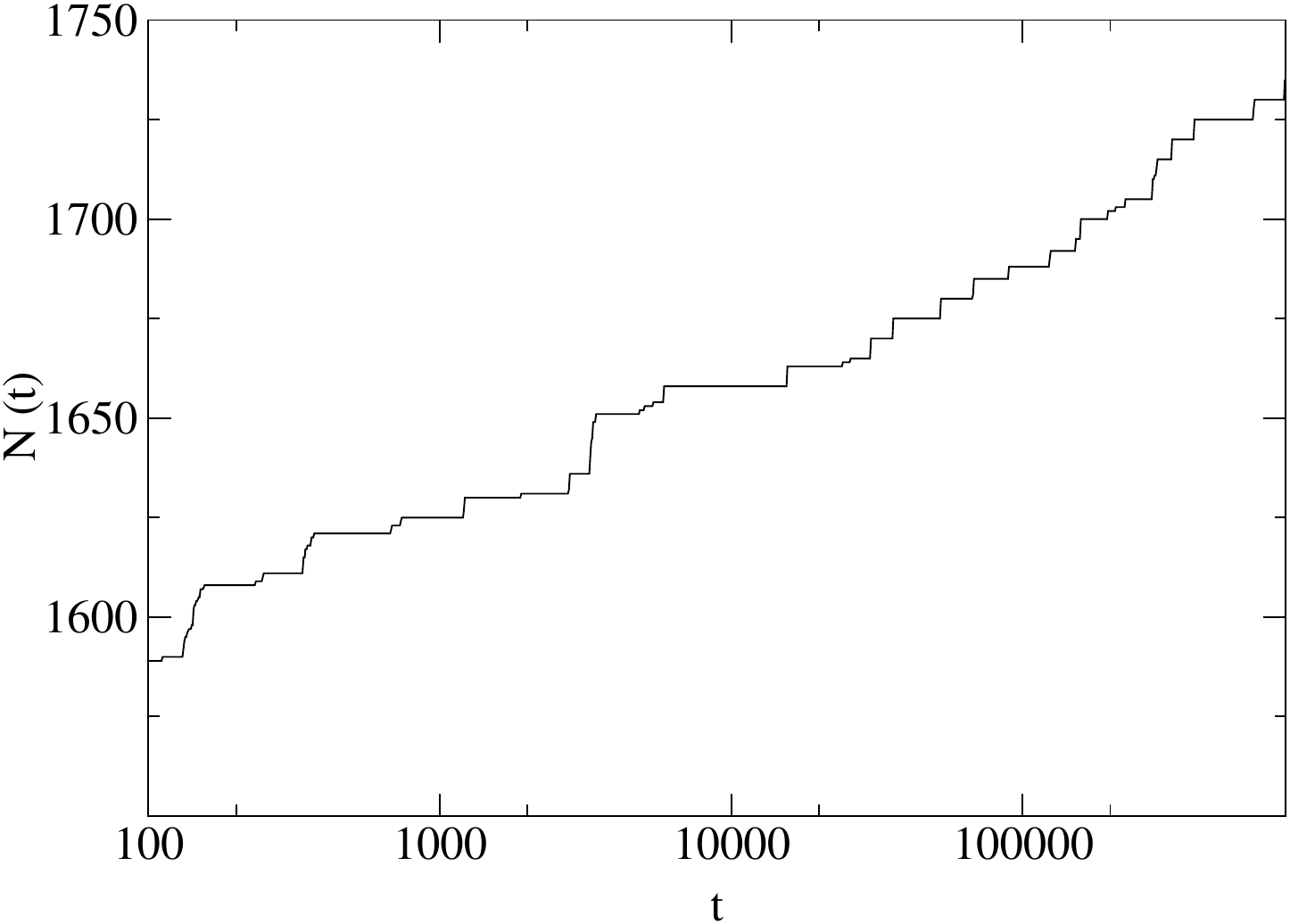}\hfill{}
\caption{The detailed time variation of the total number of vortices $N(t)$
on the system for a single realization of the pinning potential and
the thermal noise in a $8\times8\times8$ lattice for $T=0.1$. { {red} Data obtained 
simulating the ROM model briefly described in the main text.}  Notice
that $N(t)$ increases in a stepwise fashion through quakes and features
an overall logarithmic growth trend. Figure taken from Ref.~\cite{Oliveira05}.
\label{fig:steps}}
\end{figure}

The low temperature dynamical evolution of the response
to a magnetic field quickly ramped up to a constant value at $t=0$
is shown in Fig.~\ref{fig:steps}, taken from~\cite{Oliveira05}.
The total number $N(t)$ of vortices, which is plotted versus the logarithm
of time, mainly changes through nearly vertical steps which punctuate
equilibrium-like plateau values. The overwhelming majority of steps
leads irreversibly to a higher number of vortices and we hence identify
them with our quakes. 

\begin{figure}
\hfill{}\includegraphics[clip,width=8cm]{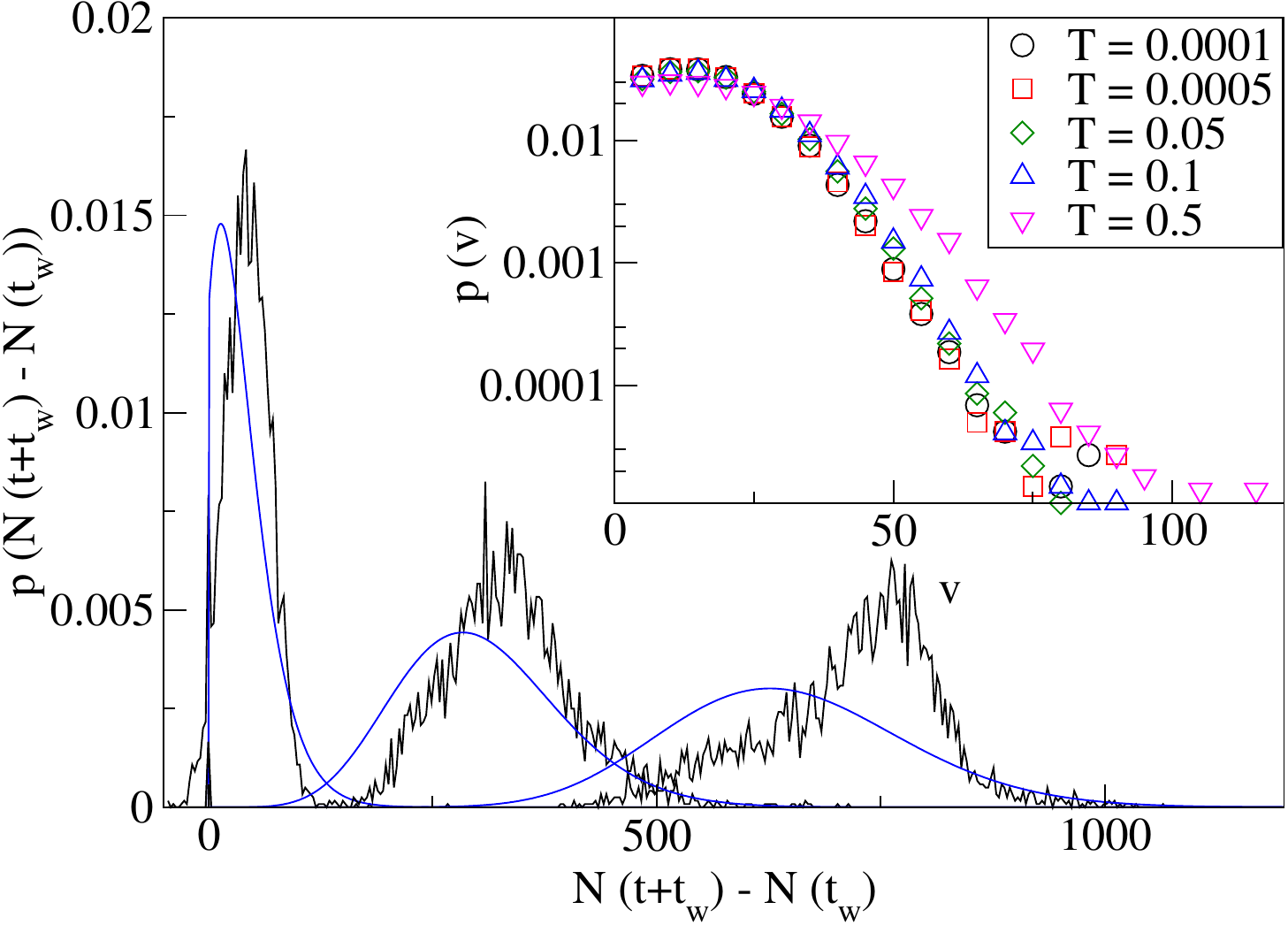} \hfill{}
\caption{(Color online) The main panel contains the temporal evolution of the
probability density function, $P(N(t+t_{w})-N(t_{w}))$, of the number
of vortices entering for $t_{w}=1000$ and three different observation
times $t=188,2791,8371$ given by the black, slightly jagged, curves.
The smooth curves (blue) are a fit to the theoretical expression (see
Eq.~(\ref{pdf_N})). The system is $16\times16\times8$ and $T=0.0001$.
The insert shows the quake size distribution for various temperatures
for the time interval between $t=1000$ and $t=10000$. For $T\le0.1$
the distribution has an approximately exponential tail. For $T=0.5$
the shape gets closer to a Gaussian.  { {red} Data obtained 
simulating the ROM model briefly described in the main text. Figure taken from Ref.~\cite{Oliveira05}.}}
\label{fig:log-Poisson}
\end{figure}

The quake size is the number $v$ of vortices it allows to enter the
system and its Probability Density Function of $v$ is shown in the
insert of Fig.~\ref{fig:log-Poisson} for several different temperatures
and three different observation times, all starting at $t_{{\rm w}}=1000$
MC sweeps. Statistical insight into the time evolution in the number
of vortices present within the system is provided by Fig.~\ref{fig:log-Poisson},
where the empirical distributions of $N(t)$ are displayed for three
different times, which are equidistantly placed on a logarithmic time
axis. The insert in Fig.~\ref{fig:log-Poisson} shows the tail of
the probability density function of the number of vortices, $p(v)$,
entering during a single quake. To a good approximation, the tail
is exponential and the time and temperature dependence of of $p(v)$
are negligible, except for the highest temperature $T=0.5$. From
Eq.\eqref{logPoisson}, the probability that exactly $q$ quakes occur
during the time interval $[t_{w},t_{w}+t]$ is log-Poisson distributed
according to 
\begin{equation}
P_{q}(t,t_{{\rm w}})==\frac{t_{{\rm w}}}{t+t_{{\rm w}}}\frac{(\alpha\log(1+t/t_{{\rm w}}))^{q}}{q!},\quad q=0,1,2,\ldots\infty,\label{logPoisson2}
\end{equation}
where $\alpha$ is the logarithmic quaking rate.

We approximate the PDF for the number of vortices $v$ which enter
during a given quake (see insert Fig.~\ref{fig:log-Poisson}) by an
exponential distribution $p(v)=exp(-v/\bar{v})/\bar{v}$, and assume
that consecutive quakes are statistically independent. The number
of vortices entering during exactly $q$ quakes is then a sum of exponentially
distributed. independent variables, and is hence Gamma distributed.
Finally the PDF of the total number of vortices entering during $[t_{w},t_{w}+t]$
by averaging the Gamma distribution for $q$ quakes over Eq.~\eqref{logPoisson2}.
This leads to the following expression for the PDF of the total number
of vortices $\Delta N=N(t+t_{w})-N(t_{w})$ entering during the time
interval $[t_{w},t_{w}+t]$ 
\begin{equation}
p(\Delta N,t)=e^{-\frac{\Delta N}{\bar{v}}-\langle q\rangle}\sqrt{\frac{\langle q\rangle}{\bar{v}\Delta N}}I_{1}\biggl(2\sqrt{\frac{\langle q\rangle\Delta N}{\bar{v}}}\biggr),\label{pdf_N}
\end{equation}
where $I_{1}$ denotes the modified Bessel function of order $1$.
To estimate $\langle q\rangle$, $\alpha=22.6$ is used for the logarithmic
quaking rate. The average $\bar{v}$ either obtained from the distributions
in the insert of Fig.~\ref{fig:log-Poisson} or from fitting Eq.~(\ref{pdf_N})
to the simulated data in the main frame of the same figure is estimated
to be $\bar{v}=16$. We note that $\bar{v}$ is essentially temperature
independent for temperatures below $T\approx0.1$. The probability
density function, $P(N(t+t_{w})-N(t_{w}))$ is plotted as a smooth
line for $t_{w}=1000$ and three different observation times $t=188,2791,8371$
in the main panel of Fig.~\ref{fig:log-Poisson}, while the jagged
line depicts the simulation results.The agreement is reasonable, considering
the numerous approximations entering the derivation.

If we imagine that individual flux lines are trapped into local energy
wells and escape them as a result of thermal fluctuations over a single
energy barrier $\Delta E$, the rate of the process would contain
an Arrhenius factor $\exp(-\Delta E/(k_{b}T))$ and feature a strong
dependence on the temperature $T$. This naive scenario is disproven
by experimental observations showing that the magnetic relaxation
rate, or creep rate, only varies weakly in a broad range of temperature,
see~\cite{Civale90}.

\begin{figure}[ptb]
\hfill{}\includegraphics[clip,width=8cm]{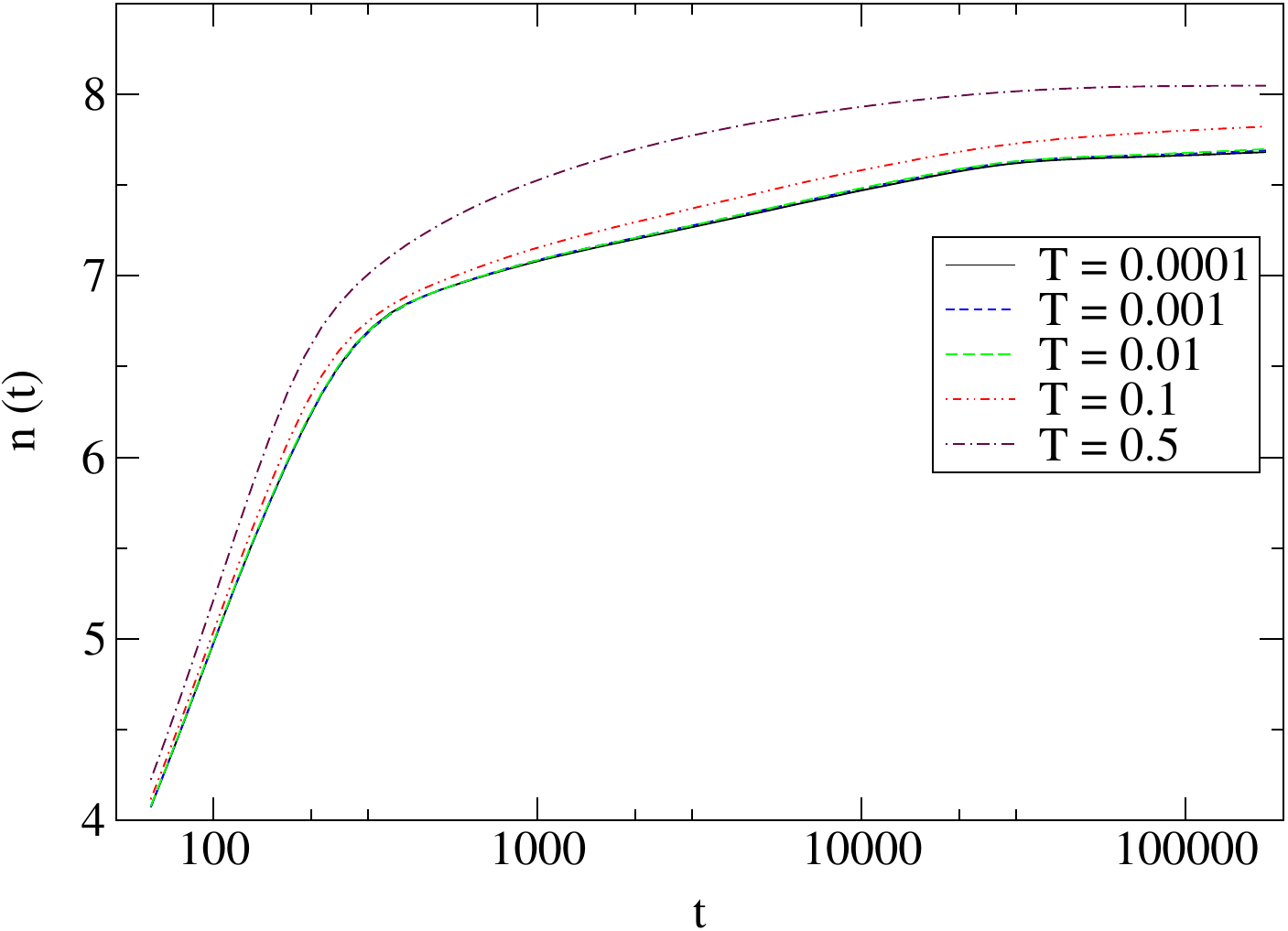} \hfill{}
\caption{(Color online) Average vortex density is plotted vs. the logarithm
of time for various temperatures. For $T\le0.1$ the vortex number
is a piecewise linear function of $\log(t)$. For $T=0.5$ the system
relaxation becomes faster. { {red} Data obtained 
simulating the ROM model briefly described in the main text. Figure taken from Ref.~\cite{Oliveira05}}.}
\label{fig:N(log(t))}
\end{figure}

To study ROM behavior in this respect, the external field is rapidly
increased at $t=0$ to a constant value. The time dependence of the
total number of vortices in the sample is then studied for several
values of the temperature $T$. The procedure is repeated for many
realizations in order to estimate an average vortex bulk density $n(t)$.
As shown in Fig.~\ref{fig:N(log(t))}, $n(t)$ gradually increases as
vortices move in from the boundary and the value of $T$ has little
effect except at the highest temperatures. Finally, in the region
$3\;10^{2}<t<10^{4}$ $n(t)$ depends linearly on the logarithm of
time. 

The near independence of the creep rate on the temperature suggests
the presence of a continuum of barriers such that a record high energy
fluctuation is enough to trigger a quake. The behavior than follows
from the independence of record statistics on the thermal noise signal
distribution. 

\subsection{\;\; Spin glasses}

\label{SG} A general property of aging systems first noticed in spin-glasses
is that short time probes, e.g., the imposition of a high frequency
AC field under cooling at a constant rate, elicit a (pseudo)equilibrium
response, while long time probes, the same external field applied
at constant temperature, reveal the true non-equilibrium nature of
the relaxation process. Interestingly, `short' and `long' must be
understood relative to the time elapsed since the initial thermal
quench, i.e., the system age $t_{{\rm w}}$. This property alone suffices
to establish the presence of a hierarchy of dynamical barriers in
configuration space. 
\begin{figure}
\vspace{-0cm}
 \hfill{}\includegraphics[clip,width=0.5\textwidth]{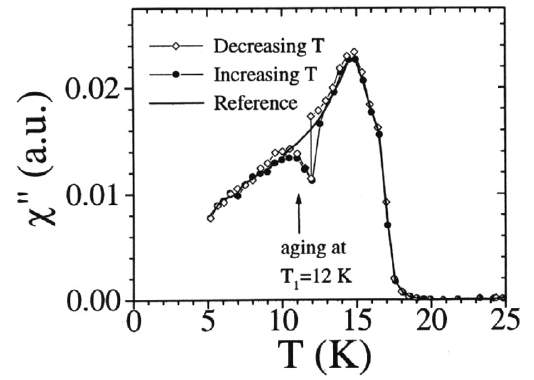}
\hfill{} \caption[Memory behaviour in spin glasses]{{\small{}Experimental results by Jonason et al.~\cite{Jonason98},
illustrating that the memory of the configurations visited by ageing
at a certain temperature is not erased by a temperature sweep.}}
\label{fig:memory} 
\end{figure}

The experiments of Jonason et al~\cite{Jonason98}, whose main result
is shown in Fig.~\ref{fig:memory}, clearly demonstrate the hierarchical
barrier structures present in spin glass configuration space. In the
figure, the imaginary part $\chi"$ of the AC susceptibility of a
spin glass is plotted as a function of temperature. Importantly, the
system is first cooled at a constant rate, except for a pause at temperature
$T=12K$, after which cooling is resumed. Once the lowest temperature
is reached, re-heating at a constant rate, without pausing at $T=12K$
is carried out. The reference curve, which results from cooling at
a constant rate, is the equilibrium result expected when the frequency
of the applied magnetic field is sufficiently high. The dip shows
that aging entrenches the system into gradually more stable configurations,
with an ensuing decrease of the susceptibility. When cooling is resumed,
the system rapidly returns to the reference curve, a so-called `rejuvenation
effect'. This indicates that the `subvalleys' explored at lower temperatures,
are all dynamically equivalent, irrespective of the aging process.
Finally, upon reheating without pause, the system remembers the configuration
space region previously explored. This `memory effect' confirms the
hierarchical nature of the energy landscape.

Thermal relaxation models associate the multi-scaled nature of aging
processes to a hierarchy of metastable components of configuration
space~\cite{Simon62,Palmer84,Hoffmann88,Sibani89}, often described
as nested `valleys' of an energy landscape as illustrated in Fig.
\ref{fig:FLandscape}. Local thermal equilibration is described in
terms of time dependent valley occupation probabilities~\cite{Sibani93},
which are controlled by transition rates over the available `passes'.
When applied to a hierarchical structure, such description gradually
coarsens over time as valleys of increasing size reach equilibrium.
That barrier crossings are connected to record values in time series
of sampled energies~\cite{Dall03,Boettcher05} is a central point
in record dynamics.

In connection with spin-glasses, RD has predictions describing Thermo-Remanent
Magnetization (TRM) data~\cite{Sibani06} where quakes are extracted
from experimental data as anomalous magnetic fluctuations. The results
described below are based on more recent numerical work~\cite{Sibani18}
to which we refer for further details, and which describes the spontaneous
magnetic fluctuations of the Edwards-Anderson model~\cite{Edwards75}.
As detailed in~\cite{Sibani18}, quakes are linked to record sized
entries in the time series of observed energy values in isothermal
simulations. We note in passing that this approach requires the use
of an event driven Monte Carlo algorithm, the Waiting Time Method~\cite{Dall03},
where event times are real numbers on scales much finer than a MC
sweep.

\begin{figure}
\hfill{}\includegraphics[bb=0bp 191.3043bp 612bp 604.522bp,clip,width=1\columnwidth]{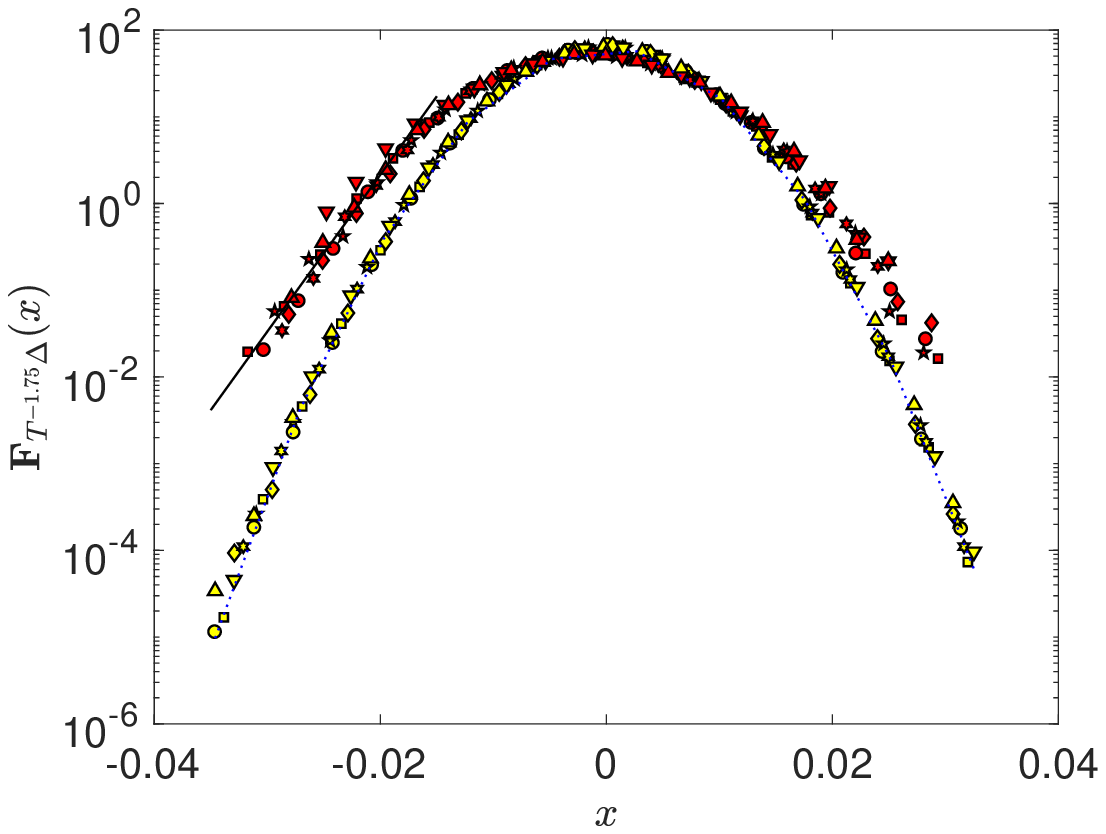}\hfill{}
\caption{Seven PDFs of energy fluctuations $\Delta$ collected at aging temperatures
$T=.3,.4,\ldots.7,.75$ and $.8$ are collapsed into a single Gaussian
PDF by the scaling $\Delta\rightarrow T^{-\alpha}\Delta,\;\alpha=1.75$,
and plotted using a logarithmic vertical scale. The data plotted with
yellow symbols are fitted by the Gaussian shown as a dotted line.
This Gaussian has average $\mu_{G}=0$ and standard deviation $\sigma_{G}\approx6.2\;10^{-3}$.
Data plotted with red symbols represent quake induced energy fluctuations
$\Delta_{{\rm q}}$ and, for negative values of the abscissa, have
estimated probabilities close to the exponential PDF shown by the
line. Figure taken  from Re.~\cite{Sibani18}}
\label{FL_stat}
\end{figure}

\begin{figure}
\hfill{}\includegraphics[bb=0bp 191.3043bp 573.75bp 589.217bp,clip,width=1\columnwidth]{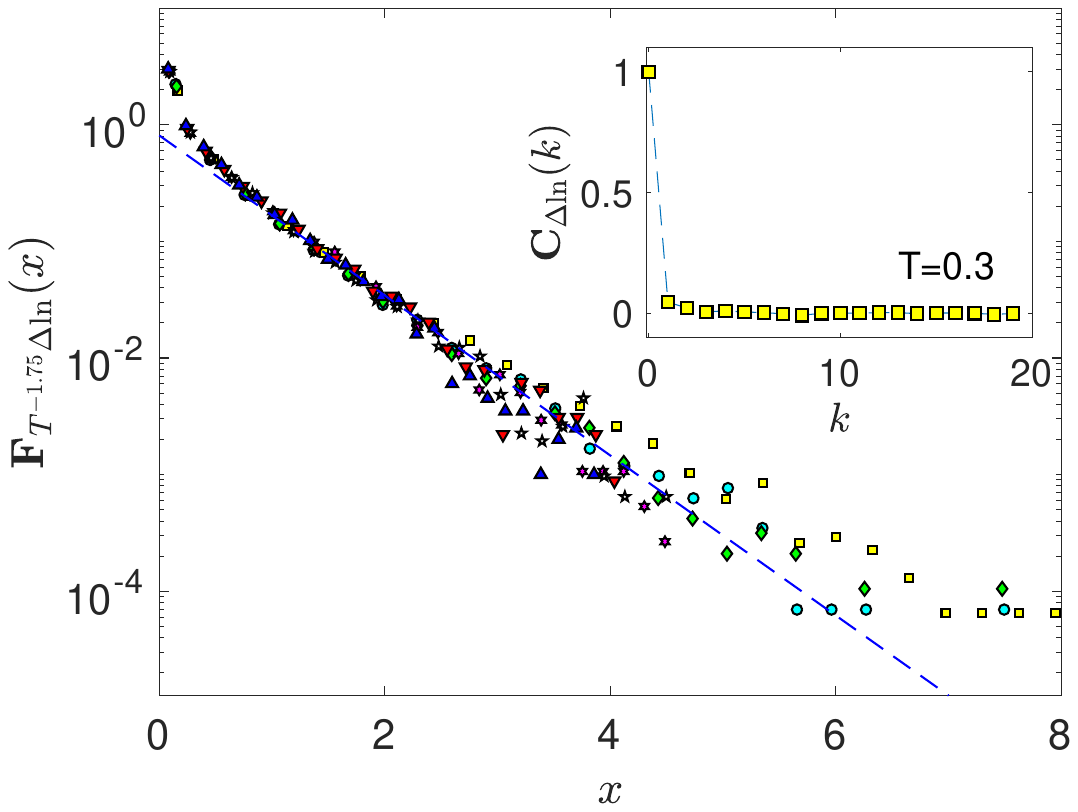}\hfill{}
\hfill{}\includegraphics[bb=0bp 191.3043bp 573.75bp 589.217bp,clip,width=1\columnwidth]{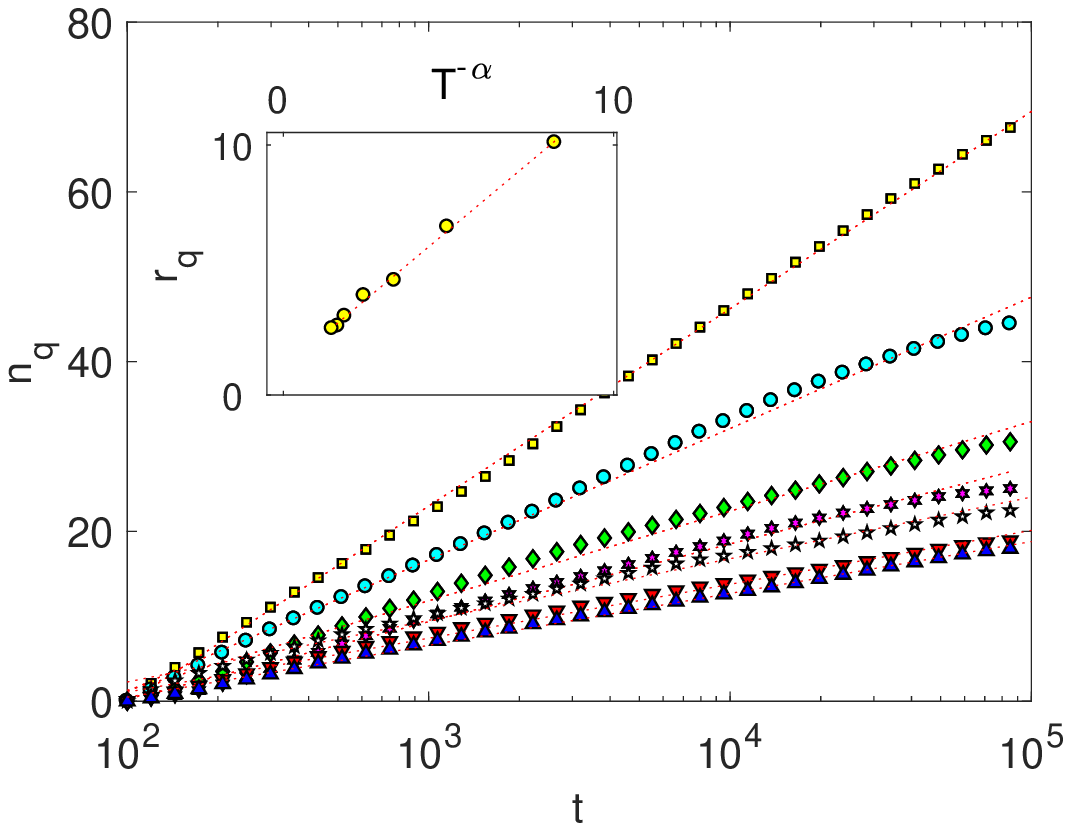}\hfill{}
\caption{Upper panel. Symbols: PDF of scaled `logarithmic waiting times' $T^{-\alpha}\Delta{\rm ln}$,
$\alpha=1.75$, for the seven aging temperatures $T=.3,.4,\ldots.7,.75$
and $.8$. Dotted line: fit to the exponential form $y(x)=.81e^{-1.57x}$.
Insert: the normalized autocorrelation function of the logarithmic
waiting times is very close to a Kronecker delta function $C_{\Delta{\rm ln}}(k)\approx\delta_{k,0}$.
The data shown are collected at $T=.3$, but similar behavior is observed
at the other investigated temperatures. Lower panel: the number of
quakes occurring up to time $t$ is plotted with a logarithmic abscissa,
for all $T$ values, with the steepest curve corresponding to the
lowest temperature. Insert: The quake rate, obtained as the logarithmic
slope of the curves shown in the main figure, is plotted vs. $T^{-\alpha}$,
where $\alpha=1.75$. The dotted line is a fit with slope $1.11$. Figure taken  from Re.~\cite{Sibani18}}
\label{DLTS}
\end{figure}

Figure.~\ref{FL_stat} shows, on a logarithmic ordinate, the PDFs
of energy changes $\Delta$ measured over time intervals of two different
types and scaled according to the temperature at which the simulation
is run. The lower curve, which is well fitted by a Gaussian of zero
mean, describes energy changes over short time intervals of equal
length. The upper curve, which feature an exponential decay on its
left wing, describes energy fluctuations observed from quake to quake,
i.e. over time intervals of varying length. Results from isothermal
simulations carried out at seven different temperatures, $T=.3,.4,\ldots.7,.75$
and $.8$ are collapsed by the scaling $\Delta\rightarrow T^{-\alpha}\Delta,\;\alpha=1.75$.
The scaling form reflects entropic effects linked to the density of
states near local energy minima.

Consider now the times of occurrence $t'$ and $t$ of two successive
quakes, $t>t'$, and form the logarithmic time difference $\Delta{\rm ln}=\ln(t)-\ln(t')=\ln(t/t')>0$,
called for short, \emph{log waiting time}. If quaking is a Poisson
process in logarithmic time, the corresponding PDF, $F_{{\Delta}{\rm ln}}(x)$
is given theoretically by 
\begin{equation}
F_{\Delta{\rm ln}}(x)=r_{q}e^{-r_{q}x},\label{quaking_r}
\end{equation}
where $r_{q}$ is the constant logarithmic quaking rate. The applicability
of equation~\eqref{quaking_r} has already been tested in a number
of different systems, including spin-glasses~\cite{Sibani07}.

The upper panel of Fig.~\ref{DLTS} shows the empirical PDFs of our
logarithmic waiting times, sampled at different temperatures and collapsed
through the same scaling as above, $\Delta{\rm ln}\rightarrow T^{-\alpha}\Delta{\rm ln}$.
The resulting PDF is fitted by the expression $F_{T^{-\alpha}\Delta{\rm ln}}(x)=.81e^{-1.57x}$,
which covers two decades of decay. Its mismatch with the correctly
normalized expression~\eqref{quaking_r} stems from the systematic
deviations from an exponential decay visible for small $x$ values.
These deviations arise in turn from quakes which occur in rapid succession
in distant parts of the system. These are uncorrelated and their statistics
is not captured by the tranformation $t\rightarrow\ln t$. The insert
shows that log-waiting times are to a good approximation $\delta$-correlated.

\begin{figure*}
\hfill{}\includegraphics[bb=0bp 325bp 762.3bp 600bp,clip,width=1\textwidth]{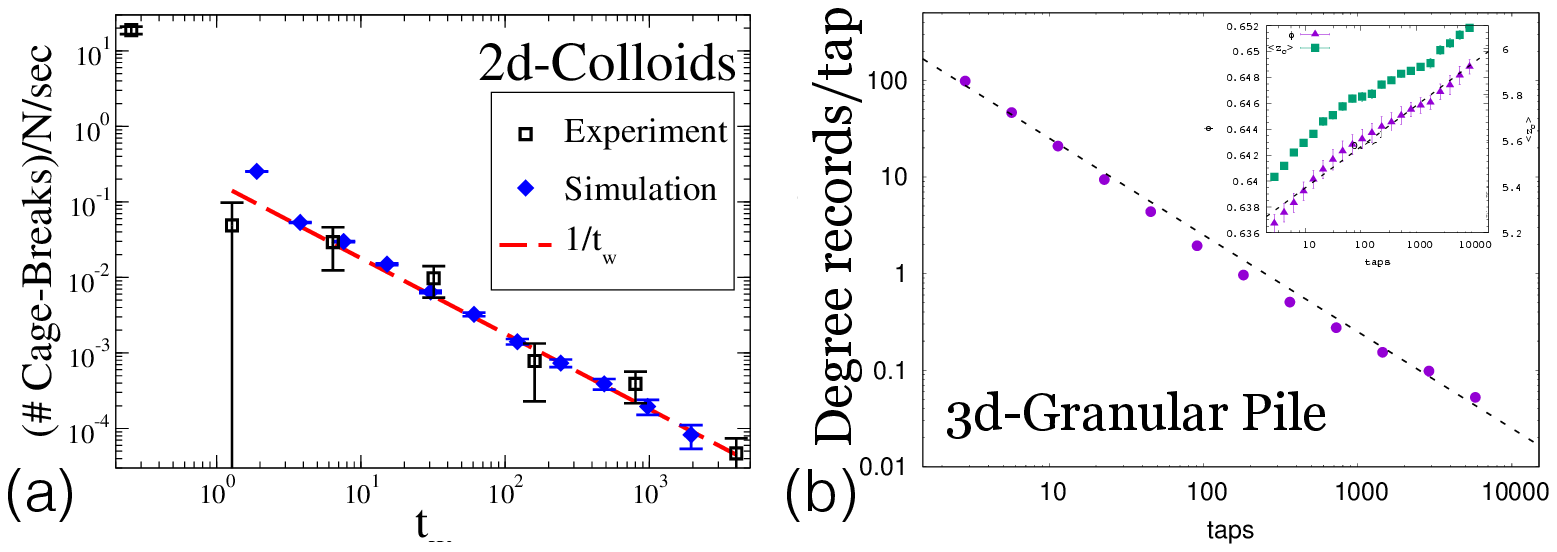}\hfill{}

\caption{{\small{}\label{fig:Decay}(a) Hyperbolic decay in the rate of intermittent
cage-breaks (``quakes'') in an experiment on jammed $2d$-colloids (black,
data from Fig.\ 2a of Ref.~\cite{Yunker09}). MD-simulation of
the colloid~\cite{Robe18} (blue) reproduces experiment over three
decades. (b) Log-Poisson statistics of aging~\cite{Boettcher18b}
in the same simulations.}}
\end{figure*}

Neglecting temporally very close events usually originating from well separated parts of the system,
 leads to the corrected number of
quakes $n_{{\rm q}}(t)$ occurring up to time $t$ which is shown
in the bottom panel of the figure for seven different aging temperatures.
The steepest curve corresponds to the lowest temperature. The red
dotted lines are linear fits of $n_{{\rm q}}(t)$ vs. $\ln t$, and
the insert shows that the logarithmic slope of the curves is well
described by the function $r_{{\rm q}}=1.11T^{-1.75}$. We note that
the logarithmic quake rate as obtained from the exponent (not the
pre-factor) of the fit $y(x)=.81e^{-1.57x}$ is $r_{{\rm q}}=1.57T^{-1.75}$.
The two procedures followed to determine the quaking rate are thus
mathematically but not numerically equivalent: in the time domain
they give the same $T^{-1.75}/t$ dependence of the quaking rate,
but with two different pre-factors. The procedure using the PDF of
the logarithmic waiting times seems preferable, due to better statistics.

Glossing over procedural difference, we write $r_{{\rm q}}=cT^{-1.75}$
where $c$ is a constant, and note that in our RD description the
number of quakes occurring in the interval $[0,t)$ is then a Poisson
process with average $\mu_{N}(t)=cT^{-\alpha}\ln(t)$. Qualitatively,
we see that lowering the temperature decreases the log-waiting times
and correspondingly increases the quaking rate. The quakes involve,
however, much smaller energy differences at lower temperatures. Considering
that $T^{-\alpha}\gg T^{-1}$, we see that the strongest dynamical
constraints are not provided by energetic barriers. As discussed in
Ref.~\cite{Sibani18}, they are entropic in nature and stem from
the dearth of available low energy states close to local energy minima.

Finally, our numerical evidence fully confirms the idea that quaking
in the Edward-Anderson model is a Poisson process whose average is
proportional to the logarithm of time. In other words, the transformation
$t\rightarrow\ln t$ renders the aging dynamics (log) time homogeneous
and permits a greatly simplified mathematical description.

\subsection{\;\; Hard sphere colloidal suspensions}

\label{HSC}

Hard sphere colloids suspensions (HSC) are a paradigmatic and intensively
investigated complex system~\cite{Weeks00,Weeks02,Courtland03,ElMasri05,Cianci06,ElMasri10,Hunter12}
with two dynamical regimes controlled by the particle volume fraction~\cite{Hunter12}:
below and above a critical volume fraction one finds a time homogeneous
diffusive regime and an aging regime, respectively. Here, the particle
mean square displacement (MSD) grows at a decelerating rate through
all experimentally accessible time scales.
\begin{figure*}
\hfill{}\includegraphics[bb=0bp 0bp 682bp 547.302bp,clip,width=0.48\textwidth]{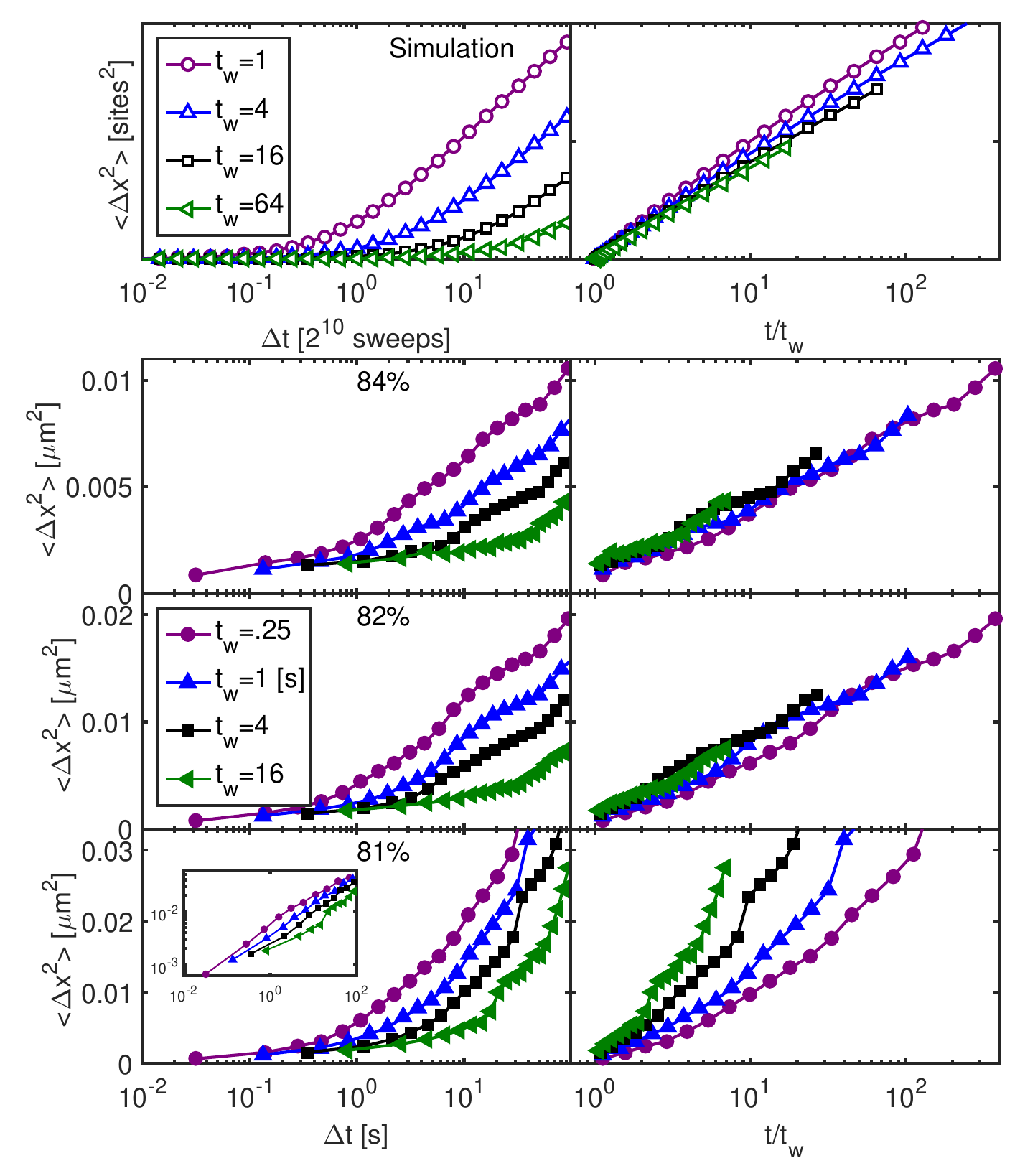}\hfill{}\includegraphics[bb=0bp 0bp 683bp 553.34bp,clip,width=0.48\textwidth]{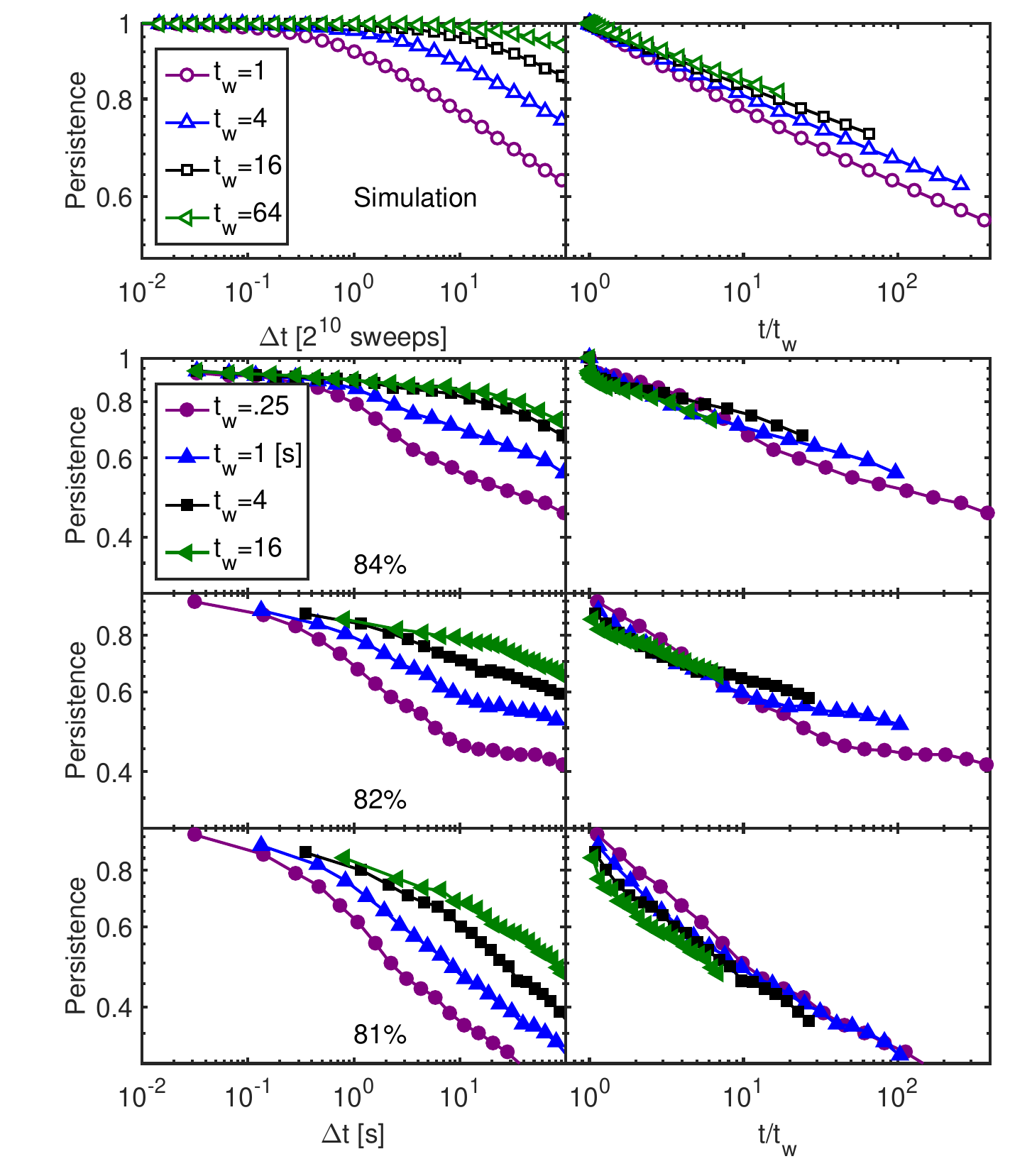}\hfill{}
\caption{{\small{}\label{fig:tiledMSD} }Mean square displacement {\small{}(MSD,
left block) and persistence (right block) in experiments of Yunker
et al~\cite{Yunker09} on 2d-colloidal suspensions at several area
fractions, re-analyzed in Ref.~\cite{Robe16}. Panels of the same row
contain the exact same data, however, each panel in the left column
of each block is plotted against the conventional lag-time $\Delta t=t-t_{{\rm w}}$,
and against the time-ratio $t/t_{{\rm w}}$ in the right column on
the log-scale suggested by RD. The experimental data are arranged
by decreasing area fractions, with $\approx84\%$, $\approx82\%$,
and $\approx81\%$, from top to bottom. The inset to the bottom left
panel contains the same MSD data as that panel, but plotted on the
log-log scale for ordinary diffusion; at 81\%, the system is no longer
jamming and full aging is avoided. Both, MSD and persistence, in the
jamming regime ($>81\%$) of the experiments closely follow the RD
predictions.}}
\end{figure*}
\begin{figure*}
\hfill{}\includegraphics[bb=0bp 376.311bp 742.5bp 594.175bp,clip,width=1\textwidth]{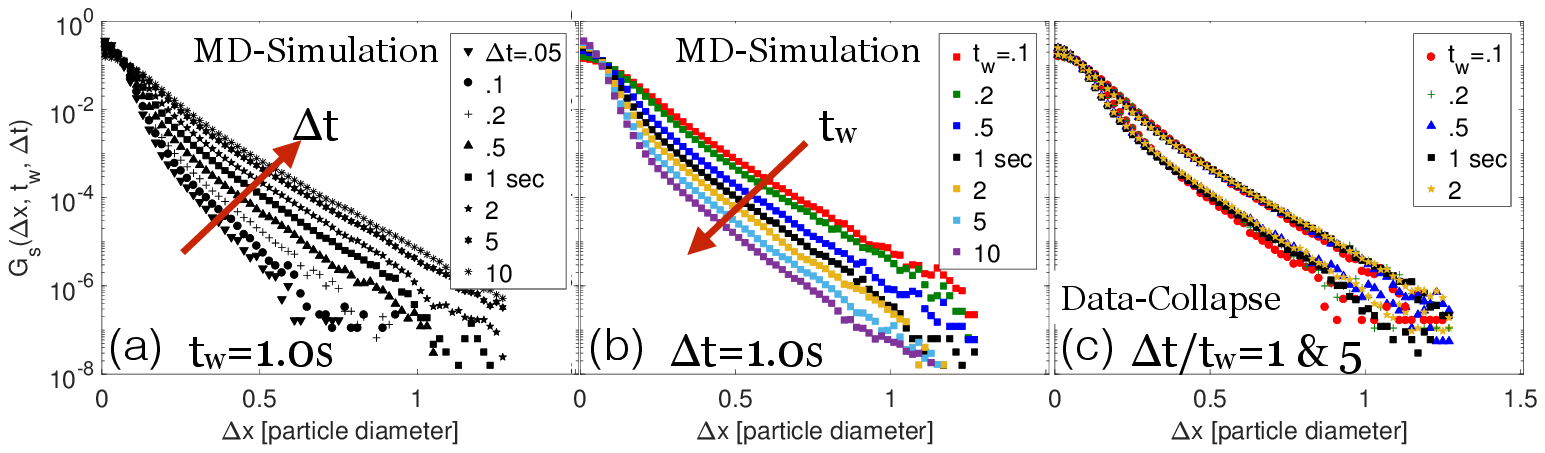}\hfill{}
\caption{{\small{}\label{fig:vanHove}Van Hove distribution of single particle
displacements $\Delta x$ over a time-window $\Delta t$ starting
at $t_{w}$ after quenching, from the MD simulation of a 2D colloid
~\cite{Robe18}. It spreads out more with increasing $\Delta t$ (a),
but less with increasing $t_{w}$ (b). Data from (a-b) collapse for
any fixed ratio of $\Delta t/t_{w}$ (c), as predicted by RD.}}
\end{figure*}

The first RD description of HSC dynamics~\cite{Boettcher11} re-analyzed
available experimental data~\cite{Courtland03}, pointing out that
the particles' mean squared displacement (MSD) grows logarithmically
with time. Furthermore, a heuristic model explaining the experimental
observations was proposed and later extensively investigated numerically~\cite{Becker14b}.
More recently Robe et al.~\cite{Robe16} highlighted the hyperbolic
decay of the rate of `intermittent' events in 2D colloidal suspensions
experiments by Yunker et al.~\cite{Yunker09}, shown in Fig.~\ref{fig:Decay}(a).
\begin{figure*}[t!]
\hfill{}\includegraphics[bb=22.95bp 176bp 543.15bp 589.217bp,clip,width=0.48\textwidth]{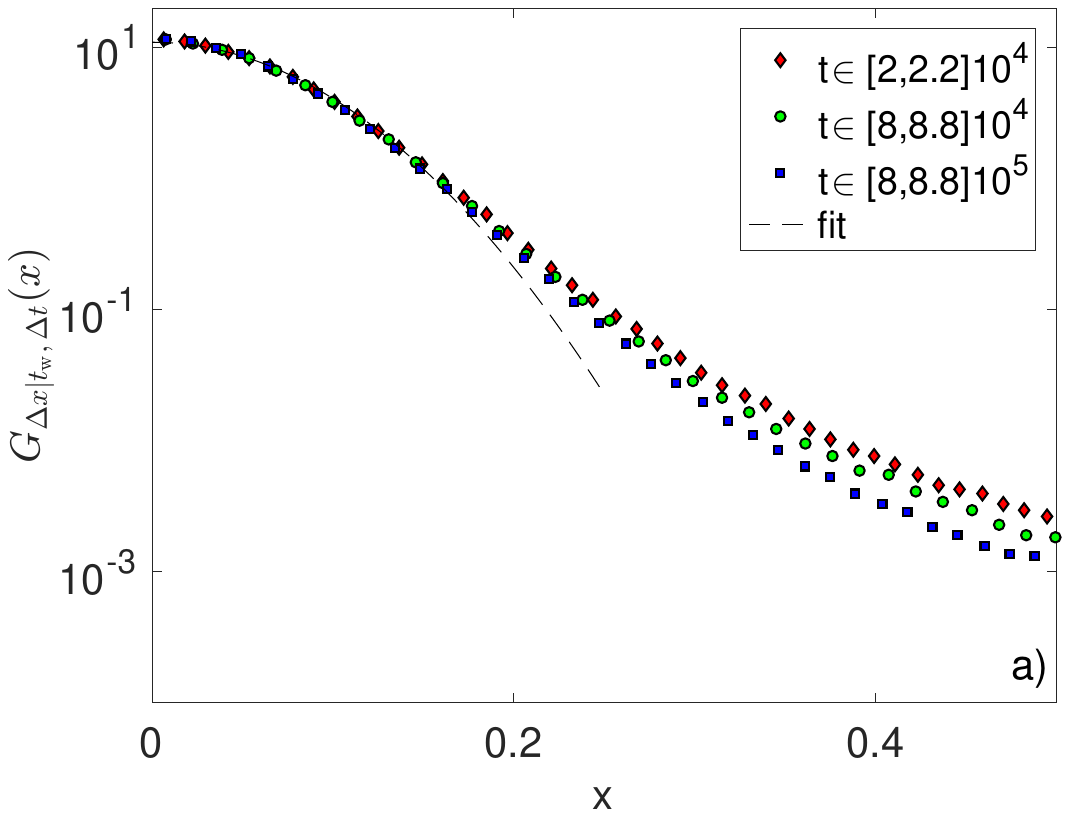}\hfill{}\includegraphics[bb=22.95bp 176bp 543.15bp 589.217bp,clip,width=0.48\textwidth]{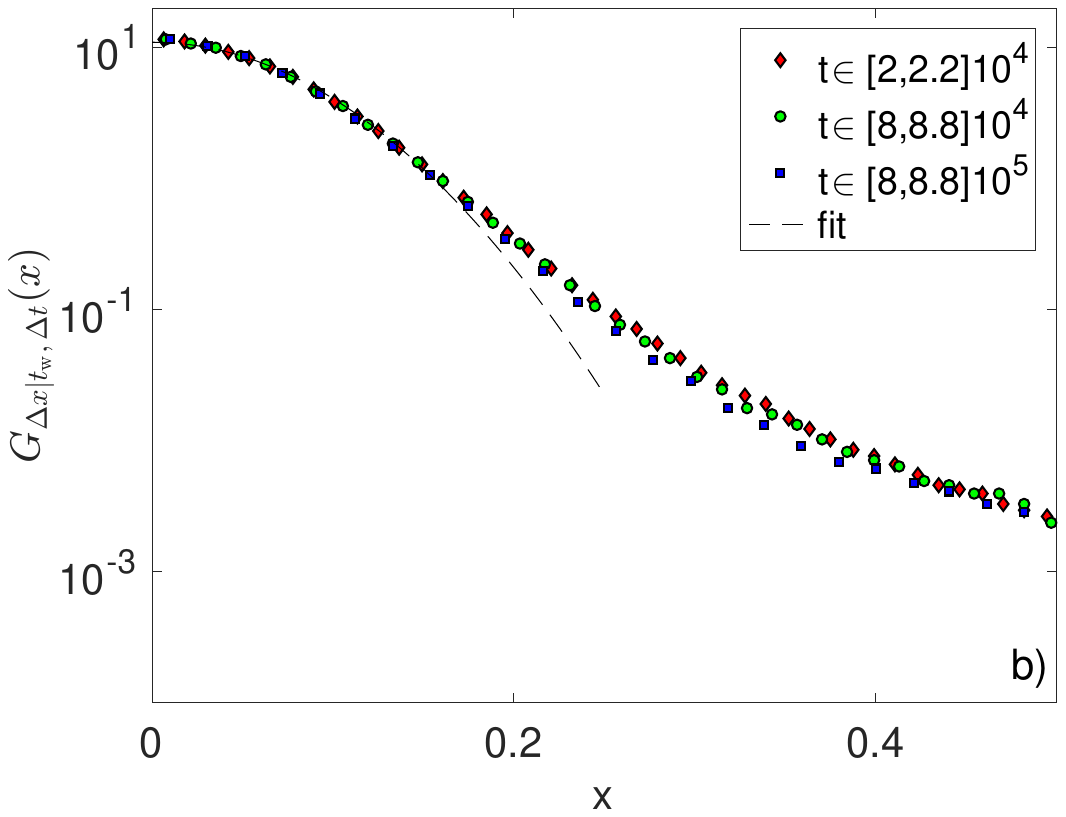}
\hfill{}\caption{For three different ages, $t_{{\rm w}}=2\cdot10^{4},8\cdot10^{4}$
and $8\cdot10^{5}$, the PDF of $\Delta x$, a one dimensional particle
displacement sampled over a time interval $\Delta t\ll t_{{\rm w}}$,
is plotted with a logarithmic ordinate. In both panels, the staggered
line is a fit to a Gaussian of mean $\mu_{G}=0$ and standard deviation
$\sigma_{G}=0.05\sigma$, where $\sigma$ is the average particle
diameter. 
Left hand panel: the same time interval $\Delta t=100\tau$ is used
for all three values of $t_{{\rm w}}$. Right hand panel: time intervals
$\Delta t=100,400,4000\tau$ growing proportional to the system age
are used. The volume fraction of this system is $\phi=0.620$. { {red} Figure taken from ~\cite{Sibani19} } }
\label{fig:intermittency}
\end{figure*}

\begin{figure*}
\hfill{}\includegraphics[bb=45.9bp 183.6522bp 550.8bp 596.87bp,clip,width=0.47\textwidth]{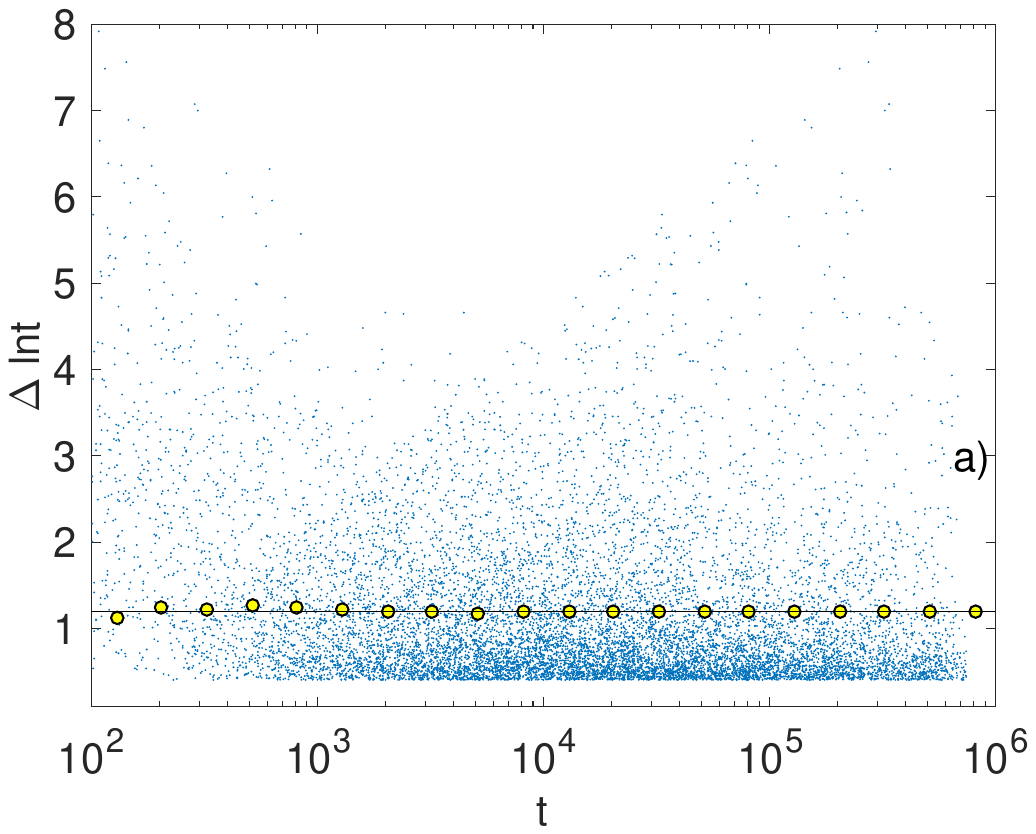}\hfill{}\includegraphics[bb=45.9bp 183.6522bp 550.8bp 596.87bp,clip,width=0.47\textwidth]{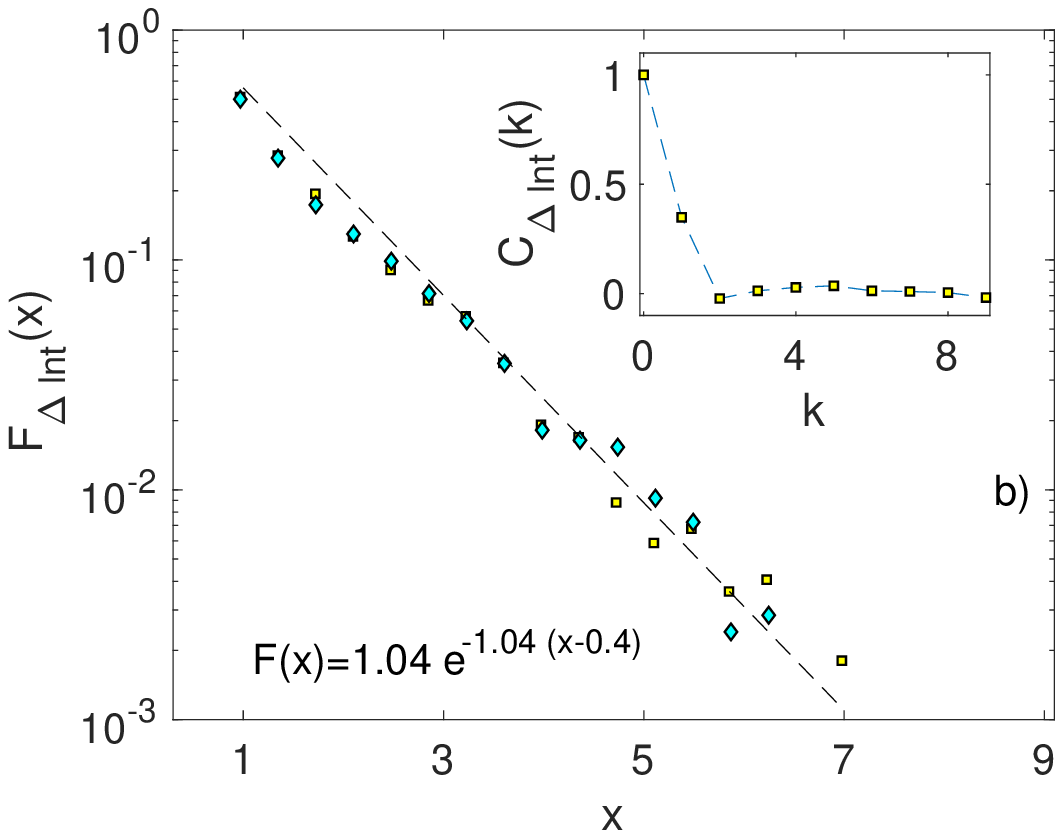}\hfill{}
\caption{a) Logarithmic waiting times $\Delta{\rm \ln t}_{k}=\ln t_{k}/t_{k-1}$,
where $t_{k}$ is the time of the $k$'th cage breaking are observed
in small domains of the simulation box and plotted, on a log scale,
vs. the time at which they are observed. Only data with $\Delta{\rm \ln t}_{k}>0.4$
are included. The yellow circles are local binned log-time averages
of these data and the line is the average logarithmic waiting time.
That local averages are nearly independent of log-time is evidence
that the transformation $t\rightarrow\ln(t)$ renders the dynamics
log-time homogeneous. b): the PDF of the `log waiting times is estimated
and plotted for two independent sets of simulational raw data, using
yellow square and cyan diamond symbols, respectively. The line is
an exponential fit to both estimated PDFs. The insert shows the normalized
autocorrelation function of the sequence of logarithmic waiting times
corresponding to the yellow square PDF. To a good approximation, the
log-waiting times are uncorrelated and their PDF decays exponentially,
which implies that quaking is a log-Poisson process. The volume fraction
of the system is $\phi=0.620$.{ {red} Figure taken from ~\cite{Sibani19} }}
\label{fig:quake_stat}
\end{figure*}

The present exposition gives an overview of a recent numerical studies~\cite{Robe18,Sibani19},
to be consulted for  technical details, where large 3D poly-disperse
or 2D bi-disperse HSC undergo extensive MD simulations. 

{ {red} To give an idea of the numerical effort involved,
the MD calculations performed in~\cite{Sibani19} calculate the 3D trajectories of $N=50000$ hard spheres
of unit mass, interacting through the potential
$$U(r_{ij})=\frac{\epsilon}{3}\left(
\frac{\sigma_i+\sigma_j}{2 r_{ij} }\right)^{36}.$$ 
In  the above, 
$\epsilon$ is the unit of energy (and temperature), $r_{ij}$ is the distance between
particles $i$ and $j$ and $\sigma_i$ is a particle's diameter. The diameters are drawn from 
the uniform distribution in an interval 
centered at the mean particle diameter $\sigma$, which is taken as the unit of length.
The unit of time, formally $\tau=\sigma\surd (m/\epsilon)$ is the time it takes an isolated  particle 
 to move its own diameter at its thermal speed.
}

The initial
state is generated by a sudden expansion of the particles' volume,
leading to volume fractions $\phi$ both below and above the critical
value. The systems' development is subsequently followed for more
than six decades in time. Two different types of particle motion are
generally found in glass-formers~\cite{Pastore14}: `in-cage rattlings',
where each particle moves reversibly within a small region bounded
by its neighbors, and `cage-breakings', where `long jumps' alter neighborhood
relations. Since cage rattlings, overwhelmingly the most frequent
events, on average do not produce net translations, particle spreading
in HSC is facilitated by the much more rare cage breakings. We thus
take those long jumps to be the spatial manifestation of quakes, and
identify the latter from the statistics of single particle displacements.
Finally, we check that quakes are a log-Poisson process.

The macroscopic effect of HSC quakes is the logarithmic growth of
the particles' mean square displacement~\cite{Boettcher11,Robe16,Sibani19},
closely resembling the experiments by Yunker et al.~\cite{Yunker09},
exhibited in Fig.~\ref{fig:tiledMSD}. The persistence data also shown
there~\cite{Robe16}, i.e., the probability that a particle does \emph{not}
experience irreversible motion between times $t_{{\rm w}}$ and $t$,
is well described by the RD arguments for the emergence of power-laws
in Sec.~\ref{subsec:-Power-laws}. The decay of persistence (or the
intermediate scattering function~\cite{Becker14b}) gives a clear picture
of the spatial heterogeneity of glassy systems: If particle motion
were to follow a regular Poisson process, the persistence would decay
exponentially in $\Delta t=t-t_{{\rm w}}$, while in a log-Poisson
process, persistence becomes an exponential in the number of activations,
$\sim\alpha\ln\left(t/t_{{\rm w}}\right)$ according to Eq.~(\ref{mean_var}),
leading to a power-law decay in $\left(t/t_{{\rm w}}\right)^{\alpha}$.

The distribution of single particle displacements, aka, the self part
of the Van Hove function, was investigated in both, Refs.~\cite{Robe18,Sibani19},
using the method introduced in Ref.~\cite{Sibani05} to describe
heat transfer in the Edwards Anderson spin-glass. The probability
density function (PDF) of displacements $\Delta x$ occurring over
a short time interval $\Delta t$ (lag time) for given values of the
system age $t_{{\rm w}}$ is written as 
\begin{equation}
G_{\Delta x|t_{{\rm w}},\Delta t}(x)=\sum_{i}\delta(x_{i}(t_{{\rm w}})-x_{i}(t_{{\rm w}}+\Delta t)-x),\label{eq:vanHove}
\end{equation}
which is normalized over all values of the dummy variable $x$. Specifically,
$G$ is sampled by collecting all positional changes $x_{i}(t_{{\rm w}}-\Delta t)-x_{i}(t_{{\rm w}})$
occurring at age $t_{{\rm w}}$ over time intervals of duration $\Delta t$,
i.e. in the interval $I(t_{{\rm w}})=(t_{{\rm w}},t_{{\rm w}}+\Delta t)$.
To improve the statistics, spherical and reflection symmetry is used
to \emph{i)} merge the independent displacements in all orthogonal
directions into a single file, representing a fictitious `x' direction,
and \emph{ii)} to invert the sign of all negative displacements. Compatibly
with the requirement $\Delta t\ll t_{{\rm w}}$ needed to associate
$G$ with a definite age $t_{{\rm w}}$, a $\Delta t$ much larger
than the mean time between collisions is preferable, as it accommodates
many in-cage rattlings. Figure~\ref{fig:intermittency} depicts PDFs
of single particle displacements in a 3D colloid of volume fraction
$\phi=0.620$~\cite{Sibani19}. Small displacements have an age independent
Gaussian PDF, corresponding to the staggered line, while larger displacements
strongly deviate from Gaussian behavior, as already seen in Fig.~6
of Ref.~\cite{ElMasri10}. The displacements occur over short time
intervals of length $\Delta t\ll t_{{\rm w}}$ and are sampled in
three longer observation intervals of the form $[t,1.1t]$ where 
 $t=2\;10^{4}\tau,\;t=8\;10^{4}\tau\;$ and $t=8\;10^{5}\tau$,
 where $\tau$ is the simulational time unit. Since the
length of these intervals is only a tenth of the time at which observations
commence, aging effects occurring during observation can be neglected
and $t$ can be identified with the system age, i.e. $t_{{\rm w}}\approx t$.
In panel a) of Fig.~\ref{fig:intermittency} a single value $\Delta t=100\tau$
was used for all data sampling, while in panel b) values proportional
to the system age, $\Delta t=100,400$ and $4000\tau$ were used.

That the central part of $G_{\Delta x|t_{{\rm w}},\Delta t}$ is a
Gaussian distribution with zero mean indicates that displacements
of small length arise from many independent and randomly oriented
contributions, which stem from multiple in-cage rattlings. The typical
size of the cage can then be identified with the standard deviation
of the Gaussian part of the PDF which is seen to be independent of
age. The exponential tail is produced by cage-breakings, i.e. displacements
well beyond the cage size. The weight of the non-Gaussian tail in
panel a) of Fig.~\ref{fig:intermittency} is seen to decrease with
increasing age while the length distribution of displacements of length
exceeding $0.5\sigma_{G}$ is shown in~\cite{Sibani19} to be exponential
and age independent. Panel b) of Fig.~\ref{fig:intermittency} shows
that scaling $\Delta t$ with the age $t_{{\rm w}}$ reasonably collapses
the data. The same effect is obtained in~\cite{Robe18} for 2D colloidal
suspensions and for other values of the ratio $\Delta t/t_{{\rm w}}$,
see Fig.~\ref{fig:vanHove}. Summarizing, the particle displacement
statistics provides a cage size estimate and a way to identify quakes
as cage breakings.

The key step of quake identification in a given setting has some leeway,
but in spatially extended systems spatio-temporal correlations play
a major role: The existence of spatial domains, a property which reflects
the strong spatial heterogeneity of glassy dynamics~\cite{Chaudhuri07},
is required in a RD description of HCS. Spatially extended aging systems
of size $N$~\cite{Sibani03} contain an extensive number $\alpha(N)\propto N$
of equivalent spatial domains. Events occurring in different domains
are statistically independent while those occurring in the same domain
have long-lived temporal correlations. These are formally removed
in RD by the transformation $t\rightarrow\ln t$. If this device works,
the total number of quakes occurring in the system between times $t_{{\rm w}}$
and $t>t_{{\rm w}}$ is a Poisson process with average 
\begin{equation}
\mu(t,t_{{\rm w}})=\alpha(N)r_{q}\ln(t/t_{{\rm w}}),\label{grandPoisson}
\end{equation}
where $r_{q}$ is the average logarithmic quake rate in each domain.
The number of quakes is extensive and grows at a constant rate in
log-time and at a rate proportional to $1/t$ in real time. Within
each domain, the rate $r_{q}$ can be read off the log-waiting time
PDF $F_{\Delta{\ln t}}(x)=r_{q}e^{-r_{q}x}$, i.e. the probability
density that the log-waiting time to the next quake equals $x$. Usually,
only a minuscule fraction of configuration space is explored during
an aging process, and many variables do not participate in any quake.
Hence, the domain size can grow in time with no changes in $\alpha(N)$,
which is best understood as the number of \emph{active} domains where
quake activity occurs. In Ref.~\cite{Sibani19}, the simulation box
is subdivided into $16^{3}$ equal and adjacent sub-volumes, each
containing, on average, slightly more than ten particles. The size
is the largest possible yielding log-time homogeneous quake statistics.
Let $t_{k}$ denote the time of occurrence of the $k$'th quake. Panel
a) of Fig.~\ref{fig:quake_stat} shows for the 3D colloid that the
log-waiting times between successive quakes occurring within the same
domain, $\Delta\ln{t}=\ln t_{k}-\ln t_{k-1}$, are uniformly distributed
on a logarithmic time axis, while panel b) shows that \emph{i)} they
are uncorrelated, which we take as a proxy for independence, and \emph{ii)}
exponentially distributed. Together these properties imply that quaking
is a log-Poisson process within each domain, and by extension in the
whole system. We emphasize that subdividing the system into domain
is crucial when calculating the log-waiting times. Similar data can
be obtained for the 2D colloid, and even in other materials, as shown
in Fig.~\ref{fig:Decay}(b). It demonstrates that, when too many events blur 
together (e.g., in a domain with too many particles $n$), it hides the local impact of the decelerating
activated events and the exponential tails weaken. The data collapses
when counts are rescaled by $n$ (see Insets). In Ref.~\cite{Boettcher18b}
we have shown identical behavior for spin glasses as well as for our heuristic model~\cite{Becker14b} of RD, 
 revealing a property renewal processes~\cite{Bouchaud92,Schulz2014,Lomholt2013}
can not reproduce without extra assumptions~\cite{Lomholt2013}.

The logarithmic growth of the particles' MSD, resembling the data
in Fig.~\ref{fig:tiledMSD}, is also measured and displayed in~\cite{Sibani19,Robe18}.
Here we simply note that the `long jumps' associated to quakes have
an age \emph{independent} PDF. Hence, each jump increments the MSD
by an age independent amount, and the MSD and the number of quakes
are simply proportional.

\section{\;\; Discussion \& Outlook}

\label{Out} Record Dynamics offers a coarse-grained statistical description
of metastable systems evolving through a series of punctuated equilibria,
including `classic' glassy systems~\cite{Sibani05,Sibani06a,Boettcher11,Robe16,Robe18,Sibani18,Sibani19}
as well as ecosystems of biological~\cite{Becker14,Andersen16,Jensen18,Sibani11a}
and cultural~\cite{Nicholson16,Arthur17} origin.

A clear distinction is made between the fluctuation dynamics within
each equilibrium state and the dynamics of the quakes which punctuate
them. RD posits that quakes are a log-Poisson process~\cite{Sibani03},
whose decelerating nature stems from the record high free energy barriers
which must be successively overcome to destroy an equilibrium state
and generate the next. This mesoscopic description can be verified,
possibly falsified, by investigating the statistics of alleged quakes
in observational and computational data.

SOC and RD are both coarse-grained statistical descriptions of stationary
respectively non-stationary dynamics of complex systems. In spite
of this important difference, they both rely on the presence of a
multitude of marginally stable metastable attractors~\cite{Tang87}.
These attractors are fixed in SOC, but gradually evolve in RD, where
quake triggering configuration changes are systematically demoted
to reversible fluctuations. This is implied by the fact that a record
barrier only counts as such at the first crossing, and reflects the
hierarchical nature of the underlying landscape.

Continuous Time Random Walks (CTRW)~\cite{Kenkre73,Sher75,Lomholt13,Schultz14}
describe metastable system in terms of jumps from trap to trap characterized
by a fixed waiting time distribution. Once a jump has occurred, the
system is back to square one and ready to jump again in the same fashion.
CTRW are thus renewal processes, which produce, as RD also does, the
sub-diffusive behavior they were designed for, if the waiting time
probability density function is chosen to be a power-law. However,
CTRW do not offer case specific microscopic justifications of this
choice.

In spite of some superficial similarities, CTRW and RD rely on orthogonal
physical pictures of complex behavior~\cite{Sibani13a}. Whether
one or the other should be chosen in a specific application must rest
on empirical evidence~\cite{Boettcher18} and, mainly, on the philosophical
inclinations of the investigator.

In the RD applications presented, the focus has been \emph{i)} to
provide the physical background, \emph{ii)} to show how the log-Poisson
nature of the quaking process can be extracted from empirical data
and \emph{iii)} how RD predicts macroscopic behavior.

One RD feature which plays a role in spatially extended system, and
which has not been highlighted so far, stems from the simple mathematical
fact that the sum of two independent Poisson processes with averages
$m_{1}$ and $m_{2}$ is itself a Poisson process with average $m_{1}+m_{2}$.
An extended system with short ranged interactions can be treated as
composed of independent parts with linear size of the order of the
correlation length. If quaking in each part is described by a log-Poisson
process, the same applies to the whole system. Consequently, the quaking
rate is an extensive quantity which scales proportionally with the
system size. This is as reasonably expected and can be advantageous
when analyzing experimental or numerical data from different sources,
since the system size dependence can be scaled out. Finally, since
the average and variance of a Poisson process are identical, the distribution
itself becomes sharp in the thermodynamic limit, which means that
RD correctly recognizes the self-averaging property of large systems.

There are a number of RD applications not discussed for space reasons,
of which one, the Parking Lot Model~\cite{Sibani16} deserves a brief
mention because \emph{i)} its barrier hierarchy is purely entropic
and \emph{ii)} it features a length scale increasing with the system
age, a real space property generally expected in evolving systems
with growing ergodic components.

In conclusion, RD connects microscopic interactions and macroscopic
phenomenology via a mesoscopic description, based on log-time homogeneity
of the quake dynamics. To analyze a new problem with RD one needs
an hypothesis, which can of course be iteratively improved, on what
quakes are and how to identify them. Once this is done the dynamical
consequences of the quakes in real as well as configuration space
can be studied. The approach is generally applicable to systems going
through metastable equilibria and invariably produces a decelerating
dynamics.

Interestingly, this is not the case for human cultural evolution,
a staged process with rapid transitions between stages, not presently
discussed, which is accelerating rather than decelerating. However,
as recently argued in~\cite{Rasmussen19}, the natural variable of
that process is not time in SI units, but the number of inter-personal
interactions a quantity growing super linearly in time. In this case,
an RD description of an underlying decelerating optimization process
can lead to a process which accelerates, when described as function
of wall-clock time.
%

\end{document}